\documentclass[aps,prb,twocolumn,shortbibliography,superscriptaddress,article]{revtex4-1}
\usepackage{epsfig}
\usepackage{epstopdf}
\usepackage{amsmath}
\usepackage{amsfonts}
\usepackage{amssymb}
\usepackage{hyperref}
\usepackage{bm}
\usepackage{makecell}
\usepackage{rotating}
\usepackage{hyperref}
\usepackage{multirow}
\usepackage{graphicx}
\usepackage{float}

\usepackage{graphicx}
\usepackage{dcolumn}
\usepackage{bm}
\usepackage{color}

\usepackage{tikz,xcolor,hyperref}

\definecolor{lime}{HTML}{A6CE39}
\DeclareRobustCommand{\orcidicon}{%
	\begin{tikzpicture}
	\draw[lime, fill=lime] (0,0)
	circle [radius=0.16]
	node[white] {{\fontfamily{qag}\selectfont \tiny ID}};
	\draw[white, fill=white] (-0.0625,0.095)
	circle [radius=0.007];
	\end{tikzpicture}
	\hspace{-2mm}
}

\foreach \x in {A, ..., Z}{%
	\expandafter\xdef\csname orcid\x\endcsname{\noexpand\href{https://orcid.org/\csname orcidauthor\x\endcsname}{\noexpand\orcidicon}}
}

\begin{document}

\title{Staggered Dzyaloshinskii-Moriya and canting angle in centrosymmetric altermagnetic \\ and ferromagnetic phases: influence on the anomalous Hall effect and Weyl points}

\author{Mathews Benny\orcidM}
\email{mbenny@magtop.ifpan.edu.pl}
\affiliation{International Research Centre Magtop, Institute of Physics, Polish Academy of Sciences, Aleja Lotnik\'ow 32/46, PL-02668 Warsaw, Poland}

\author{Xujia Gong\orcidA}
\affiliation{International Research Centre Magtop, Institute of Physics, Polish Academy of Sciences, Aleja Lotnik\'ow 32/46, PL-02668 Warsaw, Poland}

\author{Kamil Jamroszczyk}
\affiliation{International Research Centre Magtop, Institute of Physics, Polish Academy of Sciences, Aleja Lotnik\'ow 32/46, PL-02668 Warsaw, Poland}

\author{Amar Fakhredine\orcidF}
\affiliation{Institute of Physics, Polish Academy of Sciences, Aleja Lotnik\'ow 32/46, 02668 Warsaw, Poland}

\author{Giuseppe Cuono\orcidG}
\affiliation{Department of Materials Science, University of Milan-Bicocca, Via Roberto Cozzi 55, 20125 Milan, Italy}
\affiliation{Consiglio Nazionale delle Ricerche (CNR-SPIN), Unit\'a di Ricerca presso Terzi c/o Universit\'a “G. D’Annunzio”, 66100 Chieti, Italy}

\author{Rajibul Islam\orcidR}
\affiliation{Lomare Technologies Limited London Street, London EC3R 7LP, UK}

\author{Jan Skolimowski\orcidC}
\email{jskolimowski@magtop.ifpan.edu.pl}
\affiliation{International Research Centre Magtop, Institute of Physics, Polish Academy of Sciences, Aleja Lotnik\'ow 32/46, PL-02668 Warsaw, Poland}

\author{Carmine Autieri\orcidB}
\email{autieri@magtop.ifpan.edu.pl}
\affiliation{International Research Centre Magtop, Institute of Physics, Polish Academy of Sciences,
Aleja Lotnik\'ow 32/46, PL-02668 Warsaw, Poland}
\affiliation{SPIN-CNR, UOS Salerno, IT-84084 Fisciano (SA), Italy}

\date{\today}
\begin{abstract} 
We present a simple methodology to compute the anomalous Hall conductivity (AHC) as a function of the canting angles in ferromagnets and altermagnets, starting from a nonmagnetic Hamiltonian obtained from first-principles calculations that preserves the full symmetry of the crystal structure. Magnetism is introduced by including on-site spin splitting, spin-orbit coupling, and spin-canting angles. As a representative material, we study SrRuO$_3$, which supports spin canting and exhibits a sign change of the AHC.
In the ferromagnetic phase, the low-energy AHC is found to be close to zero at the Fermi level, in agreement with experimental observations. We show that the dependence of the AHC on the relevant physical parameters is most pronounced in the central region of the electronic bandwidth. We determine the symmetry-allowed components of the AHC for different magnetic orders in the large family of transition-metal perovskite ABO$_3$ compounds with space group $62$, including the spontaneous in-plane anomalous Hall effect. Within density functional theory, we evaluate the range of spin-canting angles in SrRuO$_3$ and demonstrate that it is suppressed as electronic correlations increase. By analyzing the AHC as a function of the canting angle, we find that the collinear magnetic configurations contribute most to the AHC, while spin canting plays a secondary role in determining its magnitude in non-collinear ferromagnets and altermagnets. However, canting can become relevant and induce a sign change of the AHC when the collinear magnetic state exhibits an AHC close to zero. Finally, we investigate the locations of Weyl points in the Brillouin zone and their evolution as a function of the canting angle.
\end{abstract}


\maketitle

\section{Introduction}


Over the past decades, significant effort has been devoted to understanding the effects of a net Dzyaloshinskii-Moriya interaction (DMI) on the magnetotransport properties of magnetic materials, since it promotes spin canting and can lead to the formation of chiral magnetic textures such as skyrmions, especially in ferromagnetic systems\cite{RevModPhys.97.031001,Bogdanov2020,Jena2021,guo2017spin}. 
On the other hand, researchers have investigated the effect of \emph{staggered} DMI in altermagnets\cite{839n-rckn,PhysRevB.111.054442}. DMI is a relativistic, antisymmetric component of the exchange interaction arising from first-order in the spin-orbit coupling (SOC). SOC preserves time-reversal symmetry and therefore is unable to generate magnetism or weak ferromagnetism in systems with Kramers degeneracy. Consequently, in centrosymmetric and time-reversal-symmetric systems, SOC alone is insufficient to lift spin degeneracy or induce a net magnetic moment. In altermagnetic compounds, however, time-reversal symmetry is broken while the net magnetization remains zero in the non-relativistic limit\cite{Smejkal22beyond,Smejkal22}. In this case, the presence of an altermagnetic spin splitting allows SOC to generate a finite weak ferromagnetic moment. The altermagnetic spin-splitting is therefore a necessary condition for the emergence of weak ferromagnetism in centrosymmetric systems. However, also higher orders in SOC with a behavior similar to the DMI can produce weak ferromagnetism\cite{PhysRevLett.132.176702,839n-rckn} 
When weak ferromagnetism is symmetry-allowed, the anomalous Hall conductivity (AHC) is likewise symmetry-allowed.~\cite{doi:10.1126/sciadv.aaz8809,PhysRevLett.130.036702}. The AHC effect can be further enhanced by the presence of Weyl points near the Fermi energy~\cite{10.1063/5.0158271}. The orientation of the Hall vector is strongly dependent on the N\'eel vector, which is defined as the difference between the magnetization vectors of the two inequivalent magnetic sublattices~\cite{turek2022altermagnetism,shao2023neel}. 
It has been predicted that the anomalous Hall response can also serve as a sensitive probe of the N\'eel vector~\cite{PhysRevLett.124.067203,Fang2024}. Recently, the link between anomalous Hall conductivity and Fermi surface geometry has been emphasized.\cite{Derunova2025}
A more general discussion of the symmetry-allowed terms obtained from a multipole analysis was done for the anomalous and planar Hall effect~\cite{PhysRevX.15.031006,XIAO2025109872}.

In contrast, much less attention has been paid to the role of a \emph{staggered} DMI in ferromagnetic systems that preserve inversion symmetry\cite{dosSantosDias2023}. The staggered DMI is absent in PT-symmetric systems, but it induces spin canting in both altermagnets and ferromagnets. While the staggered DMI tends to generate weak ferromagnetism in altermagnets, it tends to generate spin canting in ferromagnets, which, in the simplest cases, gives a zero net magnetization along the subdominant components\cite{Fakhredine26}. The effect of the staggered DMI in real materials on the magnetic, transport and topological properties is usually neglected.

Members of the Ruddlesden--Popper series of Sr--ruthenate oxides~\cite{ssxp-gz9l,leon2025strainenhancedaltermagnetismca3ru2o7}, and more generally transition-metal oxides~\cite{PhysRevB.107.155126}, host rotational symmetries that generate staggered DMI. These same symmetries can also give rise to altermagnetism in systems with antiferromagnetic interactions. Consequently, transition-metal-oxide perovskites represent an extraordinary platform for realizing altermagnetic materials~\cite{Cuono23orbital,10.1063/5.0252836,leon2025strainenhancedaltermagnetismca3ru2o7}.
SrRuO$_3$ is one of the most extensively studied itinerant ferromagnetic oxides~\cite{PhysRevB.91.205116}, owing to the wide tunability of its properties in thin films~\cite{RevModPhys.84.253} and at interfaces~\cite{Vanthiel21,Groenendijk20,doi:10.1021/acsanm.9b01918,lu2025insulatortometaltransitionmagneticreconstruction,D1MH01385H,Kimbell2022}. This tunability has enabled applications in quantum electronics~\cite{10.1063/5.0100912}, freestanding membranes~\cite{Lu2022}, and the investigation of curvilinear magnetism~\cite{https://doi.org/10.1002/advs.202522085}.
Using first-principles methods, several studies have explored the effects of strain~\cite{PhysRevB.77.214410,Tian2021}, electronic correlations~\cite{PhysRevB.90.165130}, reduced dimensionality~\cite{PhysRevB.90.125109,Autieri_2016,Huang2021}, and heterostructuring~\cite{10.1063/5.0043057,Filipoiu2024} on the electronic and magnetic properties of SrRuO$_3$. More recently, this material has also been proposed to host an unconventional orbital Hall effect~\cite{Peng2025}.
The presence of spin canting in SrRuO$_3$ was demonstrated using first-principles calculations~\cite{PhysRevB.105.245107}, and its impact on the anomalous Hall effect was subsequently analyzed in terms of chirality for the antiferromagnetic single layer of SrRuO$_3$~\cite{Kipp2021}. Recently, we reported relativistic spin-momentum locking in the ferromagnetic phases of SrRuO$_3$. For the ferromagnetic phase of SrRuO$_3$ with magnetization along the $z$ axis, the relativistic spin-momentum locking consists of $d$-wave magnetism, characterized by the magnetic quadrupoles $Q_{xz}$ and $Q_{yz}$, in the subdominant spin components $S_x$ and $S_y$ respectively. The dominant component $S_z$ has been shown to exhibit $s$-wave magnetism.~\cite{Gong2026}  In (111)-orientated SrRuO$_3$, recent studies revealed that a component of the AHC was observed in the same direction as the dominant component\cite{Nishihaya2025}, highlighting the important role of subdominant components in the AHC of ferromagnets.

The AHC of SrRuO$_3$ has been shown to undergo a sign reversal when interfaced with materials possessing strong spin-orbit coupling, or when magnetism is perturbed at the interface or the surface\cite{Vanthiel21,Groenendijk20,doi:10.1021/acsanm.9b01918,lu2025insulatortometaltransitionmagneticreconstruction,D1MH01385H,Kimbell2022}. Theoretically, it has been demonstrated that even in the absence of staggered DMI, the spin canting alone can induce a sign change in the Berry curvature and consequently in the AHC \cite{Brzezicki2025}. Furthermore, the sign of the AHC can be directly related to differences in the orbital angular momentum \cite{PhysRevB.109.125305}.

In this work, we aim to establish a general framework for calculating the anomalous Hall conductivity in collinear magnetic phases of transition-metal perovskites and to investigate the influence of staggered DMI on both the AHC and on the Weyl points.
The paper is organized as follows. Section~2 presents the methodology used to construct the magnetic Hamiltonian via Wannierization of nonmagnetic first-principles calculations, which preserves the crystal symmetries with high fidelity. Section~3 reports the AHC for all magnetic orders of the perovskite ABO$_3$ and analyzes their associated symmetries. In Section~4, we report the magnitude of the spin cantings and the sign reversal of the AHC; therefore, we analyze the evolution of the Weyl-point trajectories as a function of the staggered DMI. Finally, Section~6 summarizes the main conclusions of this work.


\section{Band structure calculation and Tight-binding model in non-magnetic phase}

Performing AHC calculations using density functional theory (DFT) calculations within a relativistic framework presents two major challenges.
First, the inclusion of SOC constrains the spin orientation to the underlying crystal symmetry, making it generally impossible to tune the canting angles. This limitation complicates the study of systems where non-collinear spin configurations or weak ferromagnetism play a significant role. Second, while the Wannierization of relativistic band structures is used in the standard procedure\cite{w90-ahc}, it is often a nontrivial task, especially for large systems or complex magnetic phases. The construction of maximally localized Wannier functions (MLWFs) in the presence of SOC often yields non-symmetrized tight-binding Hamiltonians, even when the procedure is carefully optimized. Although several computational packages offer Hamiltonian symmetrization routines, these implementations typically apply only to non-relativistic cases.

\begin{figure}
    \centering
    \includegraphics[width=1\linewidth]{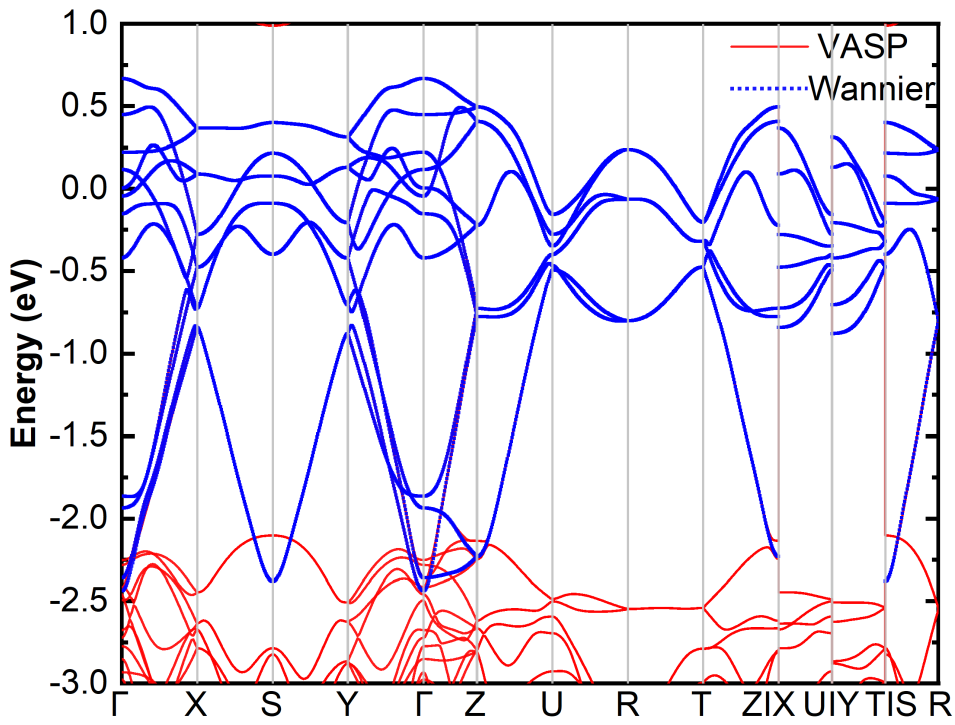}
    \caption{Comparison of the non-magnetic electronic band structures obtained from Wannier interpolation (blue dotted lines) and DFT calculations performed using the VASP code (red lines). The Wannier basis set was constructed using the $t_{2g}$ orbitals as trial wave functions.
    The Wannier-interpolated bands are in excellent agreement with the direct DFT results across the entire Brillouin zone.}
    \label{fig:NM_BS}
\end{figure}

To overcome both of these issues, we adopt an alternative procedure.
In order to tune the canting angles while preserving all symmetries of the Hamiltonian, we start from the non-magnetic (NM) Hamiltonian constructed in the Wannier-function basis. This step is generally technically simpler, as the NM Hamiltonian involves a single spin channel, which is more easily disentangled. By contrast, in the magnetic cases, the spin-splitting separates the bands into majority and minority components, increasing the likelihood of overlap with other non-magnetic bands.
The non-magnetic Hamiltonian can be written in second quantization as:
\begin{equation}
H_{NM}= \sum_{i,j} t_{i,j}^{l,m}c^{\dag}_{i,l}c_{j,m}
\label{eqnS3}
\end{equation}
where $i$ and $j$ are the indexes for the atomic sites, while $l$ and $m$ run over the Wannier basis.
The non-magnetic Hamiltonian, constructed in this way, already includes the crystal symmetries of the space group of the compounds, encompassing those symmetries that are responsible for altermagnetism. Once the local spin-splitting is added, the altermagnetic spin-splitting will appear. We add the on-site terms as the on-site spin-splitting $\Vec{h}(\theta_S,\phi_S)$ dependent on the canting angles in polar coordinates and the SOC parameter $\lambda$, which is site-dependent when there is more than one atomic species. $\theta_S$ and $\phi_S$ are the polar and azimuthal angles of the given spin. Therefore, the additional terms of the Hamiltonian read:
\begin{equation}
H_{split}+H_{SOC}= \sum_i(-\Vec{h}(\theta_S,\phi_S)_i\cdot\Vec{S}_i+\lambda_{i}\Vec{L}_i\cdot\Vec{S}_i) 
\label{eqnS4}
\end{equation}
The module of the on-site spin-splitting can be tuned and the code provides a subroutine to renormalize the Fermi level for every value of the atom-dependent input parameters. The crucial information regarding the spin evolution is included in the real space magnetic configuration $\Vec{S}_i$ and their angular dependence can be obtained by symmetry analysis 
or by studying the relativistic spin-resolved density of states (DOS)\cite{PhysRevB.111.054442,g32j-hnvz}. 
The ferromagnetic configuration with spins along the z-axis is achieved by setting $\theta_S=\phi_S=0$ for all magnetic atoms and so on for all other magnetic configurations. 
The implementation code for this research is publicly available as open-source software\cite{jskol_SOC_Code_V1_2025}. The code can be used to generate spin cantings for p, d, and f electrons on top of the non-magnetic Hamiltonian, ensuring that the final Hamiltonian possesses the correct magnetic symmetries. This code that incorporates the accuracy of first-principles and the crystal symmetries could be applied, for instance, to study magnetic multipoles \cite{yoo2025micromagneticformalismmagneticmultipoles}, non-collinear magnetic systems with commensurate magnetism\cite{Cheong2024}, non-linear Hall effect\cite{Fang2024} and magnetic topological systems\cite{https://doi.org/10.1002/adma.202201058,Brzezicki2019}.

\begin{figure}[H]
    \centering
    \includegraphics[width=1\linewidth]{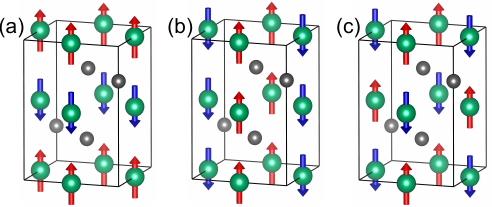}
    \caption{The altermagnetic phases of SrRuO$_3$. (a) A-type magnetic order, (b) C-type magnetic order, and (c) G-type magnetic order. The N\'eel vector is aligned along the $z$-axis in all three cases.}
    \label{fig:AFM} 
\end{figure}

\begin{figure}[H]    
    \includegraphics[width=0.99\linewidth]{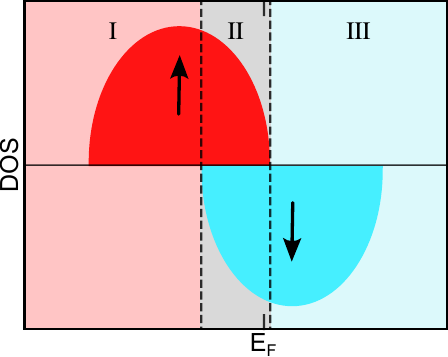}
    \caption{A depiction of the three regions observed in the band structure in which the variation of AHC with spin-splitting has considerably different behaviour. The bright red (bright blue) region denotes the DOS of spin-up (down) electrons. The three regions of the AHC, at the bottom of the bands, around the Fermi level and above the Fermi level are represented in light red, light grey and light blue, respectively. In SrRuO$_3$, the t$_{2g}$ manifold is filled to two-thirds ($2/3$).}
    \label{fig:regions}
\end{figure}

\begin{figure*}[t]
        \centering
        \includegraphics[width=\textwidth]{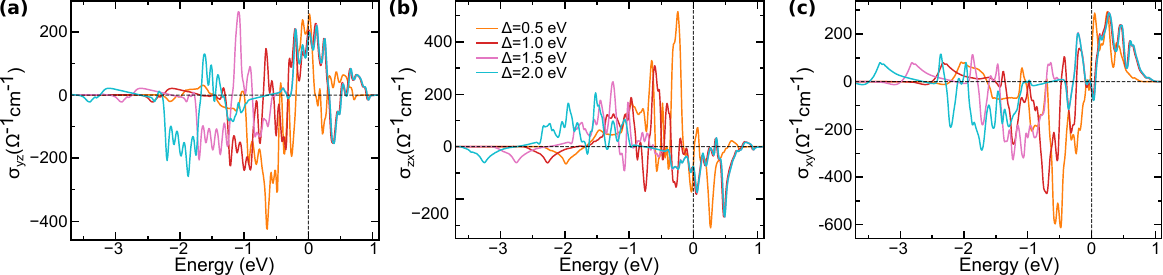}
        \caption{AHC as a function of the spin splitting for the ferromagnetic phase with magnetization along the (a,b) $x$-direction with two non-zero components: $\sigma_{yz}$ and $\sigma_{zx}$ respectively. The first is the standard AHC, while the second is the spontaneous in-plane AHC. (c) $z$-direction where the only non-zero component is $\sigma_{xy}$.}
        \label{fig:AHC_FM}
\end{figure*}

A previous version of this code was used to describe the spin-canting in EuIn$_2$As$_2$\cite{6yv6-kf97}.
For the present paper, the non-magnetic band structure of SrRuO$_3$ is represented in Fig. \ref{fig:NM_BS} with the computational details reported in Appendix A. We have wannierized the t$_{2g}$ electron bands; therefore, only Ru atoms are involved and we fix the spin-orbit to $\lambda_{Ru}$=100 meV. The value of 100 meV was extracted from the wannierization performed on the relativistic Hamiltonian and it agrees with previous literature\cite{Tamai2019-uk}. The spin splitting of ferromagnetic SrRuO$_3$ was found to be around 1 eV within Local Spin Density Approximation, as determined from the difference between the spin-up and spin-down peaks in the DOS\cite{Autieri_2016}. In this paper, we will analyze the effect of the spin-splitting in the range between 0.5 and 2.0 eV. Due to the properties of the crystal space group, the non-collinear magnetic configuration of the 4 magnetic atoms depends only on two canting parameters that, from now on, we will define as $\theta$ and $\phi$ and their physical interpretation will be given later in the paper. The parameters $\Delta$, $\theta$, and $\phi$ were tuned within a realistic energy range, and the angles were chosen in accordance with the system's symmetry.


\section{Anomalous Hall effect for the different magnetic phases as a function of the spin-splitting}

This section presents calculations of the AHC for different collinear magnetic phases as a function of the spin splitting, varied from 0.5 to 2.0 eV. We constrain these magnetic phases to be collinear, while most of them would become non-collinear after energy minimization when relativistic effects are included. All reported AHCs are spontaneous and do not require an external magnetic field. The aforementioned code\cite{jskol_SOC_Code_V1_2025} has been utilised to simulate the different collinear magnetic phases and values of the spin-splitting ($\Delta$) on top of the NM Hamiltonian. These magnetic phases include the ferromagnetic phase and three collinear altermagnetic phases depicted in Fig. \ref{fig:AFM}(a-c) with zero net magnetic moment.
It has already been shown that the different magnetic orders in perovskite ABO$_3$ compounds are altermagnetic~\cite{yuan2023degeneracy}; however, the properties of the non-relativistic spin-momentum locking depend on the specific magnetic order~\cite{Cuono23orbital}.

\begin{figure*}[t]
        \centering
        \includegraphics[width=\textwidth]{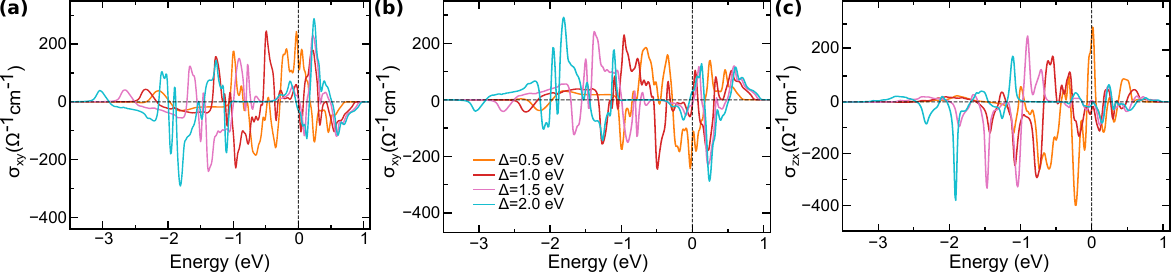}
        \caption{AHC as a function of the spin splitting for the A-type altermagnet with N\'eel vector along the (a) $x$-direction, where the only nonzero component is $\sigma_{xy}$. (b) $y$-direction where the only nonzero component is $\sigma_{xy}$. (c) $z$-direction where the only nonzero component is $\sigma_{zx}$. All magnetic configurations are constrained to be collinear and exhibit zero net magnetization.}
        \label{fig:A_type}
\end{figure*}

\subsection{Ferromagnetic order}

We begin by simulating the AHC for the ferromagnetic phase with the magnetization along the \(x\)-, \(y\)-, and \(z\)-directions as a function of the on-site spin splitting \(\Delta\).
For the ferromagnetic phase, the AHC can be qualitatively divided into three main energy regions: one above the Fermi level, one at the bottom of the bands, and one in the vicinity of the Fermi level. These three regions are schematically represented in Fig.~\ref{fig:regions}. Region I consists only of the majority states. Region II contains both majority and minority bands, while Region III consists only of minority states and can be assumed to behave similarly to Region I.

When the magnetization is along the $x$-axis, there are two contributions in the AHC tensor, namely $\sigma_{yz}$ and $\sigma_{zx}$, which are reported in Fig.~\ref{fig:AHC_FM}(a,b), respectively. 
Within DFT, we have tested the spin canting properties when the magnetization is along the $x$-axis and we found that the spins cant along the $y$-axis with two different magnitudes, resulting in a net magnetization along the $y$-axis, characteristic of a weak ferrimagnetism.
This net magnetization along the subdominant component produces an additional spontaneous AHC component, which is less common and is the magnetic analogue to the in-plane Hall effect\cite{Zhong2023}.
Therefore, the component of the AHC can be identified as a spontaneous in-plane AHC described also for other ferromagnetic compounds\cite{https://doi.org/10.1002/aelm.202500714,PhysRevB.84.104413,PhysRevX.15.031006} and for altermagnets\cite{zt4l-y18j}. 
The spontaneous in-plane anomalous Hall response was investigated in SrRuO$_3$\cite{Nishihaya2025}, but for the (111) films, which have hexagonal symmetry. The AHC exhibits a pronounced dependence on the spin splitting. As the spin splitting increases, the electronic bandwidth of the system broadens, and the AHC spans a wider energy range, particularly as states below the Fermi level are pushed further away from it. In region I, the AHC peak shifts with increasing spin splitting; with the increasing of the spin splitting, the peak progressively moves away from the Fermi level. In region III, the AHC becomes independent of the spin splitting for $\Delta \geq 1.0~\mathrm{eV}$. Region II corresponds to the crossing of the spin-up and spin-down bands; this region is the most sensitive to the variations in the parameters and also coincides with the location of the Fermi level. We note that $\sigma_{zx}$, in this case, is always relatively large at the Fermi level with a value of 200 $\Omega^{-1}\text{cm}^{-1}$. When the magnetization direction is rotated by 90 degrees in the $xy$ plane, a symmetry relation is observed between the AHC along the $x$- and $y$-directions. In particular, we find that the $\sigma_{zx}$ for the magnetization along the $x$-axis is opposite to $\sigma_{zx}$ for the magnetization along the $y$-axis, while the $\sigma_{yz}$ for the magnetization along the $x$-axis is equal to $\sigma_{yz}$ for the magnetization along the $y$-axis.

When the magnetization is along the $z$-axis, a non-zero AHC is observed in the $\sigma_{xy}$ component, as shown in Fig.~\ref{fig:AHC_FM}c, where it is plotted as a function of the on-site spin splitting $\Delta$. Region III, above the Fermi level, is characterized by a positive AHC, which becomes independent of the spin splitting for $\Delta \geq 1.0~\mathrm{eV}$. Region I is characterized by a negative AHC, while region II exhibits a small value, close to zero at the Fermi level. An AHC value close to zero would favor a sign change in the AHC, consistent with experimental observations.

\begin{figure*}[t]
        \centering
        \includegraphics[width=\textwidth]{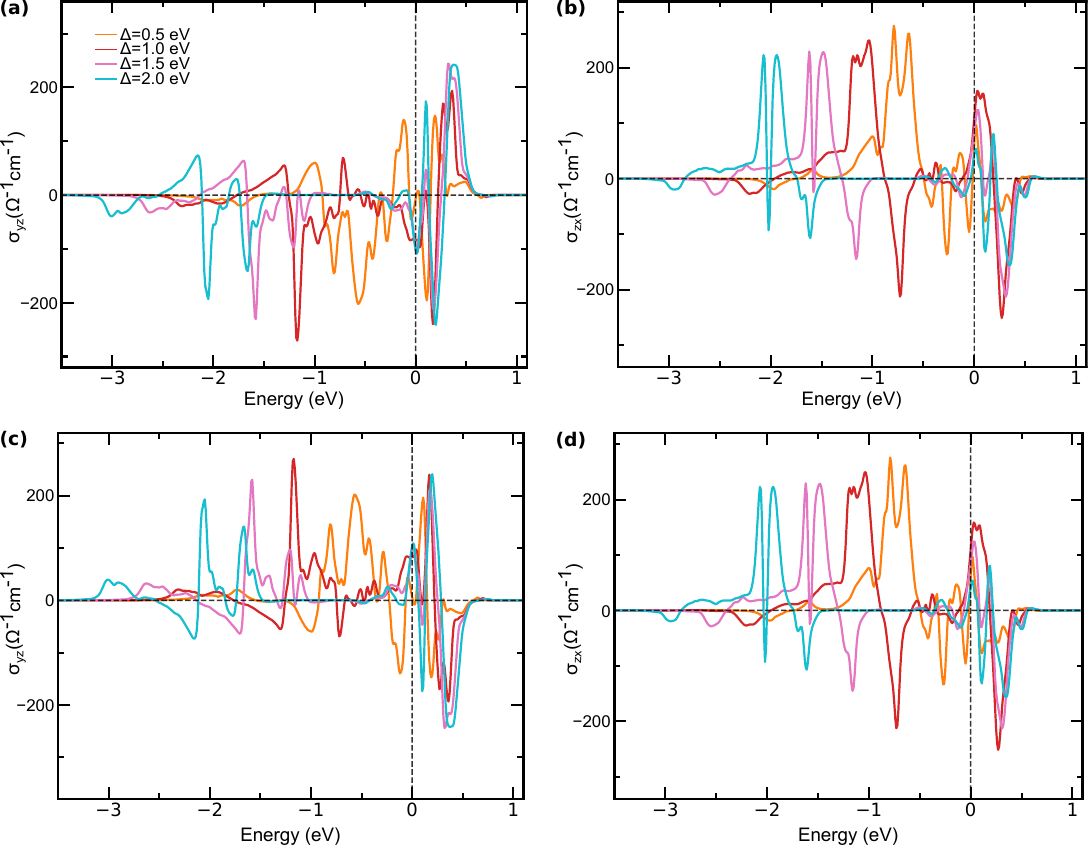}
        \caption{AHC as a function of the spin splitting for the C-type altermagnet with N\'eel vector along the (a,b) $x$-direction with two non-zero components: $\sigma_{yz}$ and $\sigma_{zx}$ respectively, (c,d) $y$-direction where there are two non-zero component: (c) $\sigma_{yz}$ and (d) $\sigma_{zx}$. All magnetic configurations are constrained to be collinear and exhibit zero net magnetization.}
        \label{fig:C_type}
\end{figure*}

\begin{figure}[h]
        \centering
        \includegraphics[width=0.99\linewidth]{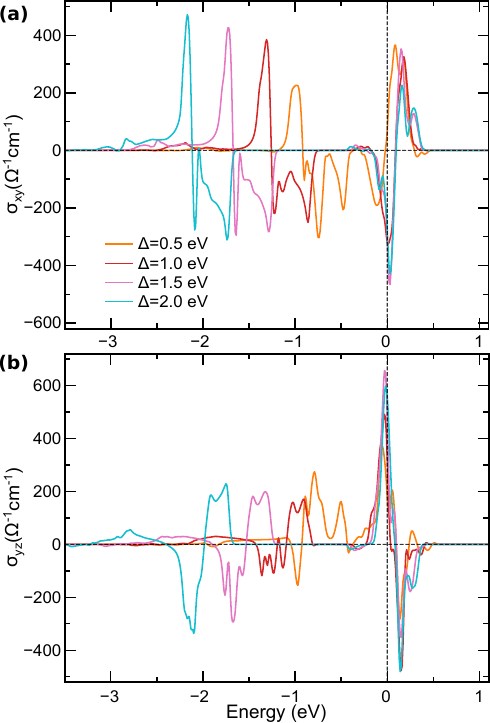}
        \caption {AHC for the G-type magnetic order with (a) N\'eel vector along both the $x$- and the $y$-directions as a function of the spin-splitting. The only nonzero component is $\sigma_{xy}$ for both cases. (b) N\'eel vector along the $z$-direction as a function of the spin-splitting. The only nonzero component is $\sigma_{yz}$. All magnetic configurations are constrained to be collinear and exhibit zero net magnetization.}
        \label{fig:G_type}
\end{figure}

\subsection{Altermagnet with A-type magnetic order}

The A-type altermagnetic phase was simulated with the N\'eel vector oriented along the $x$-, $y$-, and $z$-directions. Although weak ferromagnetism is symmetry-allowed and energetically favorable, it is explicitly constrained to be strictly zero in our calculations. Despite this constraint, the symmetry that allows weak ferromagnetism also permits a finite anomalous Hall effect along the same direction. Accordingly, a nonzero component of the AHC is obtained for all three orientations of the N\'eel vector.

For the A-type altermagnet with the N\'eel vector along the $z$-axis, the $\sigma_{zx}$ component of AHC is non-zero, depicted in Fig. \ref{fig:A_type}(c). For the other two cases, where the N\'eel vector is oriented along the $x$- and $y$-axes, respectively, the non-zero component of the AHC is $\sigma_{xy}$ in both configurations as shown in Fig. \ref{fig:A_type}(a) and Fig. \ref{fig:A_type}(b). Moreover, the AHCs in the latter two cases are the exact opposite of each other. This relation should be present due to the symmetry of the conductivity tensor\cite{PhysRevLett.126.127701}. The division into three regions that was schematically reported for the ferromagnetic phase persists slightly also in the altermagnetic phase, since it is easier to have anticrossing gaps in the middle of the bandwidth due to a larger number of bands. The quantity $\sigma_{xy}$ for these magnetic phases can reach values of ${\pm}200$ $\Omega^{-1}\text{cm}^{-1}$ at the Fermi level for $\Delta = 0.5$ $\text{eV}$.

For the altermagnetic phases, if we consider the intrinsic anomalous Hall resistivity $\rho_{xy}$ without the effect of the magnetic domains, the phenomenological formula for the anomalous Hall resistivity\cite{RevModPhys.82.1539} is not valid: 
\begin{equation}
\rho_{xy} \neq R_0 B_z + R_s \mu_0 M_z = 0
\end{equation}
since both the external magnetic field B$_z$ and the net magnetization along the $z$-axis $M_z$ are zero, while $\rho_{xy}$ (proportional to $\sigma_{xy}$) is not zero. R$_0$ and $\mu_0$ are constants described in the literature\cite{RevModPhys.82.1539}. The same is valid for the next cases that we will describe.

\subsection{Altermagnet with C-type magnetic order}

The C-type magnetic order reveals AHC contributions that are in stark contrast to those found in the other magnetic configurations. In particular, when the N\'eel vector is oriented along the $z$-direction, all components of the AHC vanish. Although time-reversal symmetry is broken in this case, the system realizes a pure altermagnetic state in which neither weak ferromagnetism nor any AHC components are symmetry allowed.
This phase constitutes the only pure altermagnetic state; nevertheless, it exhibits spin canting after energy minimization. This behavior is confirmed by DFT calculations, which show that the net magnetization remains zero despite the presence of spin canting, as in Ca$_2$RuO$_4$~\cite{Fakhredine25b}. Spin canting is therefore an intrinsic property of this class of materials, ABO$_3$ compounds with space group~62, whereas it can be symmetry-forbidden in other classes of materials\cite{Gong2026}.

In contrast, when the N\'eel vector is aligned along either the $x$- or $y$-direction, two nonzero components of the AHC emerge, namely $\sigma_{yz}$ and $\sigma_{zx}$, as shown in Fig.~\ref{fig:C_type}. The $\sigma_{zx}$ components are identical for both orientations of the N\'eel vector along the $x$- and $y$-directions. On the other hand, $\sigma_{yz}$ changes sign when the N\'eel vector is rotated from the $x$-axis to the $y$-axis. This behavior closely resembles that observed in the ferromagnetic cases with the magnetization oriented along the $x$- and $y$-directions. The presence of multiple nonzero AHC components indicates that weak ferromagnetism is symmetry allowed in both configurations, consistent with our findings for the corresponding ferromagnetic states.

\subsection{Altermagnet with G-type magnetic order}

As for the A-type magnetic order, also in the case of G-type magnetic order, a nonzero component of the AHC is obtained when the N\'eel vector is oriented along all high-symmetry directions of the lattice vectors. The G-type magnetic order with the N\'eel vector oriented along the $z$-direction exhibits a non-zero component of the AHC, namely $\sigma_{yz}$, shown in Fig. \ref{fig:G_type}c. When the N\'eel vector is oriented along the $x$- and $y$-directions, the $\sigma_{xy}$ component of the AHC is non-zero. Moreover, the AHC is identical for both the latter cases. This can be observed in Fig. \ref{fig:G_type}(a) and \ref{fig:G_type}(b).
We note that the same AHC component for the G-type magnetic order was calculated for SrRuO$_3$ thin films\cite{Samanta2020-vj}.

We observe that, for the G-type magnetic order, the AHC reaches the largest absolute values with pronounced peaks close to the Fermi level of the order of about 450~$\Omega^{-1}\text{cm}^{-1}$ when the N\'eel vector lies in the $xy$ plane, and around 600~$\Omega^{-1}\text{cm}^{-1}$ for the out of plane orientation. The latter in particular is comparable to the largest peak of the ferromagnetic phase in Fig.~\ref{fig:AHC_FM}(c), and by far larger than the value of the ferromagnetic phase at the Fermi level. 

\begin{table}[h!]
\centering
\begin{tabular}{|c|c|c|c|}
\hline
\multirow{2}{*}{Magnetic order} & \multicolumn{3}{|c|}{Only component of the Spin} \\
\cline{2-4}
 & $S_x$ & $S_y$ & $S_z$ \\
\hline
Ferromagnet & $\sigma_{yz}, \sigma_{zx}$ & $\sigma_{yz}, \sigma_{zx}$ & $\sigma_{xy}$ \\
\hline
A-type & $\sigma_{xy}$ & $\sigma_{xy}$ & $\sigma_{zx}$ \\
\hline
C-type & $\sigma_{yz}, \sigma_{zx}$ & $\sigma_{yz}, \sigma_{zx}$ & Pure altermagnet \\
\hline
G-type & $\sigma_{xy}$ & $\sigma_{xy}$ & $\sigma_{yz}$ \\
\hline
\end{tabular}
\caption{Spontaneous AHC for the different magnetic orders of the Pbnm crystal structure of the ABO$_3$ perovskite phase with B as magnetic atom. The spin has only one component, which is $S_x$, $S_y$ or $S_z$, without spin-canting.}
\label{tab:summary}
\end{table}

We summarize our results for the anomalous Hall conductivity of SrRuO$_3$ in Table~\ref{tab:summary}. These results, obtained for SrRuO$_3$, rely crucially on the fact that it is an ABO$_3$ perovskite with space group no. 62, in which the magnetism resides on the B-site atoms. The same components of the anomalous Hall conductivity reported in Table~\ref{tab:summary} will be symmetry-allowed for systems of the same material class, such as vanadates\cite{Cuono23orbital,daghofer2025altermagneticpolaronsfatealter} and other oxides with the same space group\cite{GUO2023100991}. The reported results show that for a given magnetic order, the same components of AHCs are allowed if the dominant component is along the S$_x$ or S$_y$ axis, while when the dominant component is along the S$_z$ axis, the allowed AHCs always change with respect to the other direction. For the corresponding orientation of the AHC, spin canting is symmetry allowed, and a weak ferromagnetic moment is therefore expected to emerge in DFT calculations, aligned with the AHC. In Table ~\ref{tab:summary}, there is only one magnetic phase with pure altermagnetism. Moreover, we note that Table \ref{tab:summary} contains several repeated entities. For instance, $\sigma_{xy}$ appears 5 times while $\sigma_{yz}$ in combination with $\sigma_{zx}$ appears 4 times. This can be attributed to the magnetic space group and we will illustrate this for $\sigma_{xy}$ in the next Section.

\begin{figure}
    \centering
    \includegraphics[width=0.99\linewidth]{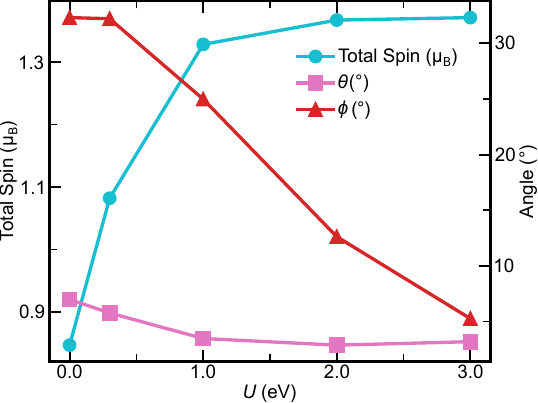}
    \caption{Evolution of the total spin $S$, spin canting $\theta$, and $\phi$ as a function of the Coulomb repulsion obtained within density functional theory.}
    \label{fig:DFT_FM}
\end{figure} 


\section{AHC sign change and Weyl point evolution in the ferromagnetic phase with spin canting}


We focus on the ferromagnetic phase with magnetization along the $z$-axis and we analyze the effect of the staggered DMI on the magnetic and transport properties. 
When SOC $\lambda$ is included, the staggered DMI, which is linear in $\lambda$\cite{PhysRev.120.91}, is implicitly taken into account. Spin canting, however, arises only after energy minimization, which can be obtained within first-principles calculations. In the first subsection, we describe and analyze the range of the canting angles obtained from first-principles calculations.

We proceed with the study of the influence of the spin canting on AHC, which has been performed only on the ferromagnetic phase with magnetization along the z-axis, since it is the experimental phase observed for bulk crystals. However, the same procedure could be applied for other magnetization directions and for the altermagnetic phases. For a realistic simulation of this compound, we will describe the behaviour of the spins S$_i$, S$_h$, S$_j$ and S$_k$ as a function of the canting angles. Since the canting angles depend on the staggered DMI parameters, by tuning the canting angles, we are manipulating the staggered DMI parameters. Therefore, the next subsections will report the influence of the staggered DMI on the anomalous Hall effect and on the Weyl points. 

\begin{figure}[h]
    \centering
    \includegraphics[width=1\linewidth]{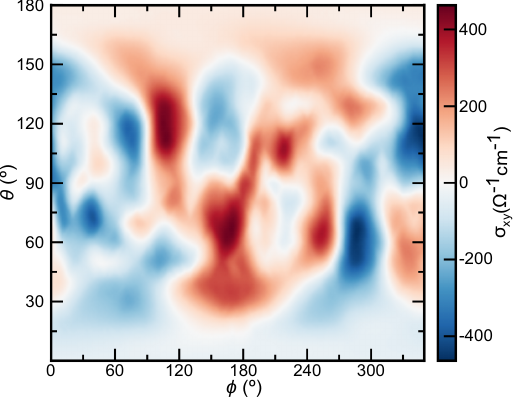}
    \caption{The variation of the AHC, $\sigma_{xy}$, of the ferromagnetic phase at the Fermi level as a function of $\theta$ and $\phi$. These data were generated for a spin-splitting of $\Delta$=1 eV.}
    \label{fig:2D_map}
\end{figure}

\begin{figure}[h]
    \centering
    \includegraphics[width=0.99\linewidth]{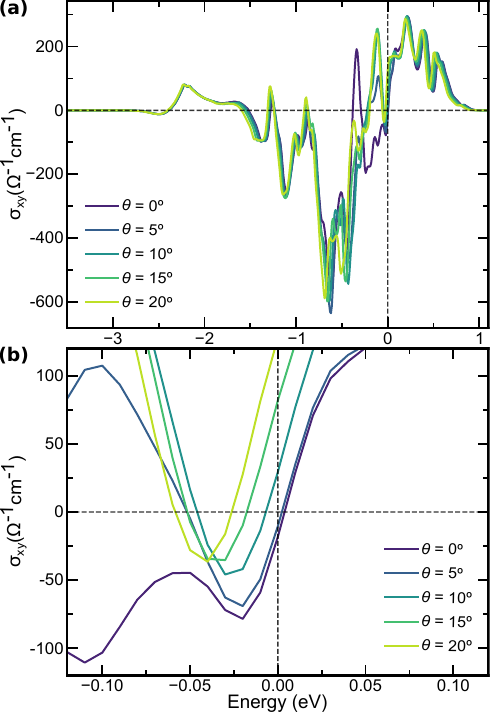}
    \caption{(a) The sign change of the AHC at the Fermi level as a function of $\theta$, keeping a constant value of $\phi$ at 0$^\circ$. (b) Magnification along the energy window between -0.12 and 0.12 eV. The value of the spin-splitting is $\Delta$=0.84 eV.}
    \label{fig:sign_change_theta}
\end{figure}

\subsection{Density functional theory for the ferromagnetic phase: $\theta$ and $\phi$ as a function of U}

We perform DFT calculations for SrRuO$_3$ by using the computational setup described in Appendix A. 
Analyzing the DFT results, we observed that in the ferromagnetic phases, the four spins in the unit cell have canting angles and follow these  equations:
\small
\begin{eqnarray} \label{Si}
S_h=(+S\sin(-\theta)\cos(-\phi),+S\sin(-\theta)\sin(-\phi),S\cos(\theta)) \\ \label{Sh}
S_i=(+S\sin(+\theta)\cos(+\phi),+S\sin(+\theta)\sin(+\phi),S\cos(\theta))  \\ \label{Sj}
S_j=(+S\sin(-\theta)\cos(+\phi),+S\sin(-\theta)\sin(+\phi),S\cos(\theta))  \\ \label{Sk}
S_k=(+S\sin(+\theta)\cos(-\phi),+S\sin(+\theta)\sin(-\phi),S\cos(\theta))
\end{eqnarray}
\normalsize
where the notation for S$_i$, S$_h$, S$_j$ and S$_k$ was obtained following the notation in the literature\cite{PhysRevB.86.094413}.
These equations define the new parameters $\theta$ and $\phi$. Our definition of the polar and azimuthal angles differs from the direction of the N\'eel vectors presented in other works on anomalous transport\cite{w52v-blqm}.
While the easy axis of the bulk SrRuO$_3$ is along the c-axis in the Pbnm notation, the staggered DMI induces spin canting along the a-axis and b-axis and creates four magnetic sublattices. From these equations, we can observe how all the subdominant components S$_x$ and S$_y$ are all different for a non-zero value of $\theta$ and $\phi$, while the dominant component S$_z$ is the same for all spins. 
Note that the set of equations (\ref{Si}-\ref{Sk}) is valid for different magnetic phases with non-zero $\sigma_{xy}$, as described in the previous Section. The previous equations represent the magnetic configuration of the magnetic space group 62.448 in the BNS setting\cite{Aroyo2006}. We can obtain the ferromagnetic phase with spins along the $z$-axis for $\theta$=$\phi$=0, G-type order with N\'eel vector along the $x$-axis for $\theta$=$\frac{\pi}{2}$ and $\phi$=0, and A-type order with N\'eel vector along the $y$-axis for $\theta$=$\frac{\pi}{2}$ and $\phi$=$\frac{\pi}{2}$. In the same way, the magnetic space group 62.448 represents the G-type order with spin component S$_z$ and nonzero $\sigma_{yz}$. For the pure altermagnet, the staggered DMI contribution to the total energy does not bring any canting; therefore, we expect that it vanishes due to symmetry considerations. The analytical form of the staggered DMI for the ferromagnetic phase of SrRuO$_3$ will be given elsewhere\cite{Benny26b}. Both ferromagnetic and altermagnetic phases can be described by the same relativistic spin–momentum locking if they belong to the same magnetic space group. The only difference is that, in one case, the dominant component is s-wave, whereas in the second case, the dominant component is d-wave.\cite{Gong2026}

\begin{figure}[h]
    \centering
    \includegraphics[width=0.99\linewidth]{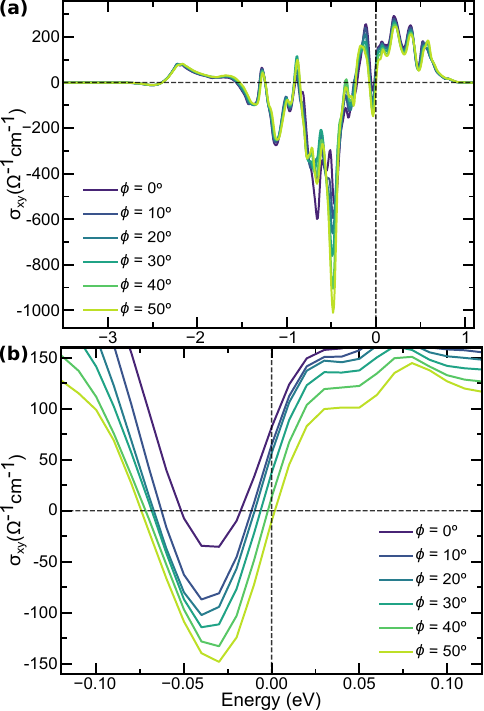}
    \caption{(a) The sign change of the AHC at the Fermi level as a function of $\phi$, keeping a constant value of $\theta$ at 15$^\circ$. (b) Magnification along the energy window between -0.12 and 0.12 eV. The value of the spin-splitting is $\Delta$=0.84 eV.}
    \label{fig:sign_change_phi}
\end{figure}


The evolution of $S$, $\theta$, $\phi$ as functions of U is reported in Fig. \ref{fig:DFT_FM}.
As observed in the DFT calculations in Fig. \ref{fig:DFT_FM}, the realistic ranges of the canting angles are 2$^\circ$ to 7$^\circ$ for $\theta$ and 5$^\circ$ to 33$^\circ$ for $\phi$. The value of $\theta$ is always smaller than the value of $\phi$. The Coulomb repulsion increases the magnetic moment, while it suppresses the spin-canting by pushing the spin-canting angles to lower values and more towards a collinear ferromagnet.

\subsection{Sign change of the AHC due to spin canting}

The spin cantings in the ferromagnetic phase induced by the DMI have been studied extensively in this section. The canting and azimuthal angles of the magnetic ground state have been varied across the entire parameter space, as shown in Fig. \ref{fig:2D_map}, for various values of spin splitting. Significant changes in the AHC at the Fermi level can be observed as the spin canting is varied. However, the variation of the AHC cannot be expressed as a simple function of the canting angles because of the numerically complicated formula of the AHC. Multiple sign changes can be observed throughout the parameter space. 

Limiting ourselves to realistic values of $\theta$ and $\phi$, we could observe sign changes in the AHC at the Fermi level corresponding to a spin-splitting of 0.84 eV. In Fig. \ref{fig:sign_change_theta}, we plot the AHC in the entire energy range between -3.5 eV and 1.1 eV and its magnification between -0.12 and +0.12 eV.  A canting angle of 10$^\circ$ was sufficient to induce a sign change in the AHC at the Fermi level, as shown in Fig. \ref{fig:sign_change_theta}. Similarly, corresponding to a canting angle of 15$^\circ$, an azimuthal angle of 50$^\circ$ leads to the sign change as shown in Fig. \ref{fig:sign_change_phi}. Therefore, not only the variation of the spin-splitting\cite{Groenendijk20}, but also the variation of the canting of the spin can induce a sign change in the AHC in SrRuO$_3$.

\begin{figure}[h]
    \centering
    \includegraphics[width=1\linewidth]{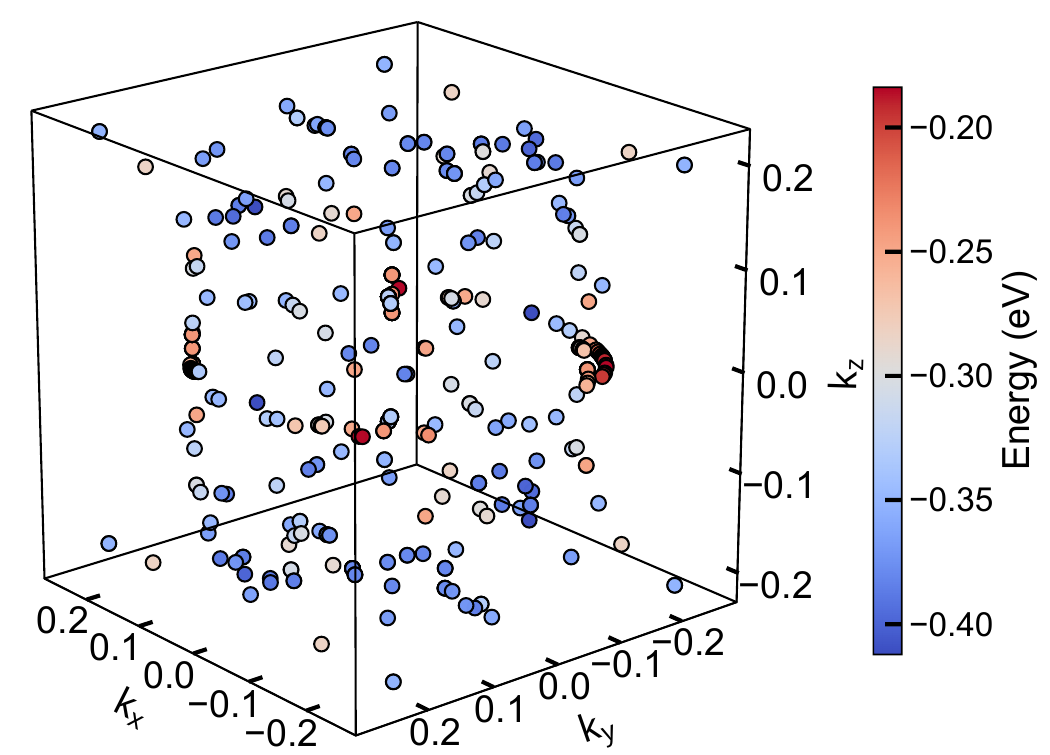}
    \caption{The nodes of the system in reciprocal space plotted as a function of their energies with respect to the Fermi level. The values of the parameters are $\Delta$=0.84 eV, $\theta$=$\phi$=0$^\circ$. The system is ferromagnetic with magnetization oriented along the $z$-axis.}
    \label{fig:all_nodes}
\end{figure}

\subsection{Evolution of the Weyl points due to spin canting}

No Weyl points have been predicted or detected at the Fermi level, but it was proposed that Weyl points physics would dominate the spin dynamics of SrRuO$_3$\cite{Itoh2016}. 
The variation of the AHC as a function of spin canting can be attributed to the evolution of the Weyl nodes induced by spin canting. To investigate this, we detect the Weyl nodes and analyze the trajectories of the Weyl points in momentum space as a function of $\theta$ and $\phi$. The Weyl nodes of the ferromagnetic phase, characterized by a spin splitting of 0.84~eV, are shown in Fig.~\ref{fig:all_nodes}. Most of the Weyl points are located in the energy window between $-0.4$ and $-0.2$~eV, corresponding to region~II defined in Fig.~\ref{fig:regions}. We observe an extended distribution of several Weyl points, with a denser accumulation at $k_z = 0$ along a circle-shaped curve.

\begin{figure}[h]
    \centering
    \includegraphics[width=1\linewidth]{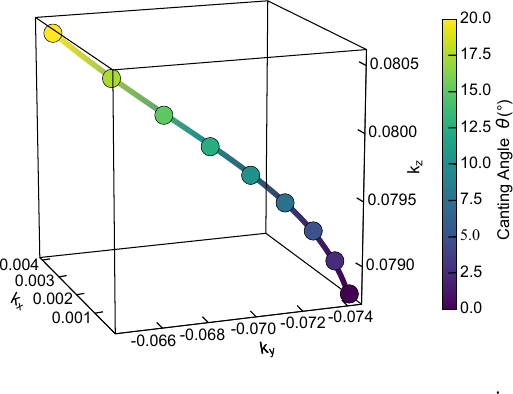}
    \caption{Evolution of a Weyl point in the system as a function of $\theta$, the other parameters are $\Delta$=0.84 eV and $\phi$=0$^\circ$. The energy of the Weyl point illustrated here is -0.35 eV. The canting induces a non-uniform variation in its energy, resulting in an energy range between -0.35 eV and -0.41 eV for the different values of $\theta$. The trajectory of the Weyl point is plotted with a solid line.}
    \label{fig:theta_nodes}
\end{figure}

\begin{figure}[h]
    \centering
    \includegraphics[width=1\linewidth]{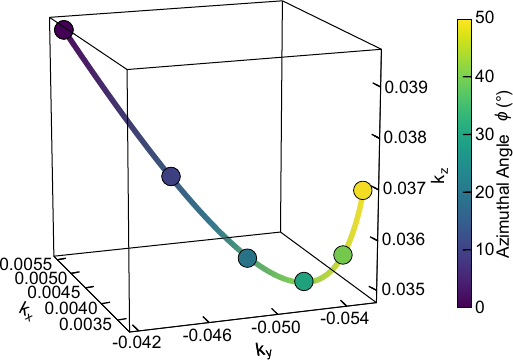}
    \caption{Evolution of a Weyl point in the system as a function of $\phi$ for $\theta$=15$^\circ$ and $\Delta$=0.84 eV. The energy of the Weyl point illustrated here is -0.38 eV. The canting induces a non-uniform variation in its energy, resulting in an energy range between -0.378 eV and -0.386 eV for the different values of $\phi$. The trajectory of the Weyl point is plotted with a solid line.}
    \label{fig:phi_nodes}
\end{figure}

The evolution of the Weyl nodes as a function of the canting angles $\theta$ and $\phi$ is presented in Figs.~\ref{fig:theta_nodes} and \ref{fig:phi_nodes}, respectively. We plot a limited region of the Brillouin zone to focus on the trajectory of a single Weyl node. The energy of the Weyl points varies by approximately 0.06~eV under changes in $\theta$, whereas it changes by only about 0.01~eV under variations in $\phi$, confirming that the Weyl points and consequently, the AHC are more sensitive to the polar angle. In both cases, the nodes exhibit a non-linear dependence on the canting angles. All the Weyl nodes numerically extracted from the band structure evolve in a similar manner to those depicted in Figs.~\ref{fig:theta_nodes} and \ref{fig:phi_nodes}. Nevertheless, the trajectories traced by the Weyl nodes as the spin canting is varied are node dependent. 


\section{Discussion and Conclusions}

We have introduced a simple and symmetry-preserving framework to compute the AHC as a function of spin-canting angles in ferromagnets and altermagnets. Starting from a nonmagnetic Hamiltonian obtained from first-principles calculations, magnetism is incorporated through on-site spin splitting, spin-orbit coupling, and controlled spin canting. Using the non-collinear ferromagnet SrRuO$_3$ as a representative system, we demonstrate that this approach captures both spin canting and the experimentally observed sign change of the AHC. We report the AHC for all directions of the spin for the ferromagnetic and altermagnetic phases. The magnetic transport properties are strongly anisotropic for all magnetic phases of SrRuO$_3$. In the ferromagnetic phase of SrRuO$_3$ with magnetization along the $z$-axis, the low-energy AHC is found to be close to zero at the Fermi level, in agreement with experiments. The sensitivity of the AHC to the on-site spin splitting is shown to be maximal in the central region of the electronic bandwidth. We further identify the symmetry-allowed AHC tensor components for different magnetic orders in transition-metal perovskite ABO$_3$ compounds crystallizing in space group 62. For the ferromagnetic phase and for the C-type order with magnetization in the $xy$ plane, both $\sigma_{zx}$ and $\sigma_{yz}$ are allowed. For the ferromagnetic phase with the dominant component in the $xy$ plane, this is the spontaneous in-plane anomalous Hall effect and we tested through DFT calculations that the subdominant component produces a weak ferrimagnetism by spin canting. We also identify the symmetry relations between different components of the conductivity tensor. 

Within DFT, we determine the range of spin-canting angles in SrRuO$_3$ and show that increasing electronic correlations suppresses canting. Our analysis of the AHC as a function of canting angles reveals that collinear magnetic configurations provide the dominant contribution to the AHC, while spin canting generally plays a secondary role in both ferromagnets and altermagnets. Nevertheless, canting can become significant when the collinear state exhibits an AHC close to zero, in which case it may induce a sign reversal. As a result, spin canting is one of the factors that must be taken into account when analyzing the sign change of the AHC in SrRuO$_3$, and more generally in other compounds that allow spin canting. Finally, we track the evolution of Weyl points in the Brillouin zone as a function of the canting angle, elucidating their role in shaping the AHC. The Weyl points and the AHC are more sensitive to the variation of the polar angle.

The methodology introduced here can also be applied to more complex magnetic structures, such as those present in Kagome lattices, as well as to the investigation of topological systems and non-linear Hall effects. In these contexts, it offers a versatile framework for capturing the interplay between lattice geometry, electronic topology, and magnetic order, and may provide new insights into emergent phenomena driven by Berry curvature and symmetry breaking.

\section*{Acknowledgments}

We acknowledge C. Ortix, R. Sattigeri, W. Brzezicki and M. Cuoco for useful discussions.
This research was supported by the "MagTop" project (FENG.02.01-IP.05-0028/23) carried out within the "International Research Agendas" programme of the Foundation for Polish Science, co-financed by the
European Union under the European Funds for Smart Economy 2021-2027 (FENG). C.A. and G.C. acknowledge support from PNRR MUR project PE0000023-NQSTI. We further acknowledge access to the computing facilities of the Interdisciplinary Center of Modeling at the University of Warsaw, Grants g91-1418, g91-1419, g96-1808, g96-1809, g103-2540, g104-2571, g104-2572 and g104-2573 for the availability of high-performance computing resources and support. We acknowledge the access to the computing facilities of the Poznan Supercomputing and Networking Center, Grants No. pl0267-01, pl0365-01, pl0471-01 and pl0694-01.

\appendix

\section{Computational details}

First-principles calculations were performed using the Vienna Ab~initio Simulation Package (VASP)~\cite{kresse1993ab,kresse1996efficiency}, 
within the framework of density functional theory (DFT) and employing the projector augmented-wave (PAW) method~\cite{kresse1999ultrasoft}. 
The exchange--correlation potential was described using the generalized gradient approximation (GGA) with the Perdew--Burke--Ernzerhof (PBE) functional~\cite{perdew1996generalized}. 
A plane-wave cutoff energy of 400~eV was used, and the total energy convergence criterion was set to $10^{-6}$~eV. 
To account for electron-electron correlations among localized 4$d$ orbitals, the DFT+$U$ formalism~\cite{liechtenstein1995density} was adopted to extract the range of the spin-canting angles. The Coulomb repulsion was tuned from 0 to 3 eV, which is a reasonable range for the 4d Ru atoms\cite{Autieri_2016,PhysRevB.90.165130,PhysRevB.90.125109}. 

SrRuO$_3$ crystallizes in an orthorhombic structure with space group $Pbnm$ (No.~62). The experimental lattice constants of SrRuO$_3$ in Pbnm notation are a=5.5670~{\AA}, b=5.5304~{\AA} and c=7.8446~{\AA}, and in Pnma they are a=5.5304~{\AA}, b=7.8446~{\AA} and c=5.5670~{\AA}.
An effective Hubbard parameter of $U = 3$~eV was employed, with Hund’s exchange coupling set to $J_H = 0.15U$. 
A Monkhorst--Pack $k$-point grid of $10 \times 10 \times 7$ was used for spin-resolved band-structure calculations. 

To extract the non-magnetic Hamiltonian, we utilize the WANNIER90 software, which transforms Bloch states into Wannier states\cite{Mostofi:2008_CPC,w90}. The AHC was calculated by utilising WannierTools\cite{wanniertools}. A $101 \times 101 \times 101$ $k$-point grid was used to perform these calculations on the Wannier Hamiltonian: convergence tests were performed with $201 \times 201 \times 201$ $k$-point grid, but minor changes were observed in the conductivity. The Weyl point positions were obtained by applying the node-finding routine provided in WannierTools.\\

\bibliography{references}

@article{kresse1993ab,
  title={Ab initio molecular dynamics for liquid metals},
  author={Kresse, Georg and Hafner, J{\"u}rgen},
  journal={Physical review B},
  volume={47},
  number={1},
  pages={558},
  year={1993},
  publisher={APS}}

@article{kresse1996efficiency,
  title={Efficiency of ab-initio total energy calculations for metals and semiconductors using a plane-wave basis set},
  author={Kresse, Georg and Furthm{\"u}ller, J{\"u}rgen},
  journal={Computational materials science},
  volume={6},
  number={1},
  pages={15--50},
  year={1996},
  publisher={Elsevier}
}

@article{zt4l-y18j,
  title = {Almost half-quantized planar Hall effects in $X$-wave magnets with $X=p,d,f,g,i$},
  author = {Ezawa, Motohiko},
  journal = {Phys. Rev. B},
  volume = {112},
  issue = {23},
  pages = {235307},
  numpages = {8},
  year = {2025},
  month = {Dec},
  publisher = {American Physical Society},
  doi = {10.1103/zt4l-y18j},
  url = {https://link.aps.org/doi/10.1103/zt4l-y18j}
}

@article{Zhong2023,
  title = {Recent progress on the planar Hall effect in quantum materials},
  volume = {32},
  ISSN = {1674-1056},
  url = {http://dx.doi.org/10.1088/1674-1056/acb91a},
  DOI = {10.1088/1674-1056/acb91a},
  number = {4},
  journal = {Chinese Physics B},
  publisher = {IOP Publishing},
  author = {Zhong,  Jingyuan and Zhuang,  Jincheng and Du,  Yi},
  year = {2023},
  month = apr,
  pages = {047203}
}

@article{RevModPhys.84.253,
  title = {Structure, physical properties, and applications of ${\mathrm{SrRuO}}_{3}$ thin films},
  author = {Koster, Gertjan and Klein, Lior and Siemons, Wolter and Rijnders, Guus and Dodge, J. Steven and Eom, Chang-Beom and Blank, Dave H. A. and Beasley, Malcolm R.},
  journal = {Rev. Mod. Phys.},
  volume = {84},
  issue = {1},
  pages = {253--298},
  numpages = {0},
  year = {2012},
  month = {Mar},
  publisher = {American Physical Society},
  doi = {10.1103/RevModPhys.84.253},
  url = {https://link.aps.org/doi/10.1103/RevModPhys.84.253}
}

@article{https://doi.org/10.1002/advs.202522085,
author = {Gao, Lei and Wang, Yuqian and Lyu, Xiangyu and Liu, Pengyu and Zhu, Mingtong and Liu, Jin and Li, Mengcheng and Ji, Ailing and Zhang, Qinghua and Gu, Lin and Ma, Libo and Cao, Zexian and Lu, Nianpeng},
title = {Strain-Assembled Crystalline SrRuO3 Microtube and Emergent Curvilinear Magnetism},
journal = {Advanced Science},
volume = {n/a},
number = {n/a},
 year = {2026},
pages = {e22085},
doi = {https://doi.org/10.1002/advs.202522085},
url = {https://advanced.onlinelibrary.wiley.com/doi/abs/10.1002/advs.202522085}
}

@article{Bogdanov2020,
  title = {Physical foundations and basic properties of magnetic skyrmions},
  volume = {2},
  ISSN = {2522-5820},
  url = {http://dx.doi.org/10.1038/s42254-020-0203-7},
  DOI = {10.1038/s42254-020-0203-7},
  number = {9},
  journal = {Nature Reviews Physics},
  publisher = {Springer Science and Business Media LLC},
  author = {Bogdanov,  Alexei N. and Panagopoulos,  Christos},
  year = {2020},
  month = jul,
  pages = {492–498}
}

@article{RevModPhys.97.031001,
  title = {Colloquium: Quantum properties and functionalities of magnetic skyrmions},
  author = {Petrovi\ifmmode \acute{c}\else \'{c}\fi{}, Alexander P. and Psaroudaki, Christina and Fischer, Peter and Garst, Markus and Panagopoulos, Christos},
  journal = {Rev. Mod. Phys.},
  volume = {97},
  issue = {3},
  pages = {031001},
  numpages = {31},
  year = {2025},
  month = {Jul},
  publisher = {American Physical Society},
  doi = {10.1103/RevModPhys.97.031001},
  url = {https://link.aps.org/doi/10.1103/RevModPhys.97.031001}
}

@ARTICLE{Tamai2019-uk,
  title     = "High-resolution photoemission on {Sr2RuO4} reveals
               correlation-enhanced effective spin-orbit coupling and
               dominantly local self-energies",
  author    = "Tamai, A and Zingl, M and Rozbicki, E and Cappelli, E and
               Ricc{\`o}, S and de la Torre, A and McKeown Walker, S and Bruno,
               F Y and King, P D C and Meevasana, W and Shi, M and Radovi{\'c},
               M and Plumb, N C and Gibbs, A S and Mackenzie, A P and Berthod,
               C and Strand, H U R and Kim, M and Georges, A and Baumberger, F",
  journal   = "Phys. Rev. X.",
  publisher = "American Physical Society (APS)",
  volume    =  9,
  number    =  2,
  month     =  jun,
  year      =  2019,
  copyright = "https://creativecommons.org/licenses/by/4.0/",
  language  = "en"
}

@article{g32j-hnvz,
  title = {Dominant orbital magnetization in the prototypical altermagnet MnTe},
  author = {Chen Ye, Chao and Tenzin, Karma and S\l{}awi\ifmmode \acute{n}\else \'{n}\fi{}ska, Jagoda and Autieri, Carmine},
  journal = {Phys. Rev. B},
  volume = {113},
  issue = {1},
  pages = {014413},
  numpages = {9},
  year = {2026},
  month = {Jan},
  publisher = {American Physical Society},
  doi = {10.1103/g32j-hnvz},
  url = {https://link.aps.org/doi/10.1103/g32j-hnvz}
}

@article{kresse1999ultrasoft,
  title={From ultrasoft pseudopotentials to the projector augmented-wave method},
  author={Kresse, Georg and Joubert, Daniel},
  journal={Physical review b},
  volume={59},
  number={3},
  pages={1758},
  year={1999},
  publisher={APS}
}

@article{liechtenstein1995density,
  title={Density-functional theory and strong interactions: Orbital ordering in Mott-Hubbard insulators},
  author={Liechtenstein, AI and Anisimov, Vladimir I and Zaanen, Jan},
  journal={Physical Review B},
  volume={52},
  number={8},
  pages={R5467},
  year={1995},
  publisher={APS}
}

@article{leon2025strainenhancedaltermagnetismca3ru2o7,
  title = {Hybrid d/p-wave altermagnetism in Ca3Ru2O7 and strain-controlled spin splitting},
  volume = {10},
  ISSN = {2397-4648},
  url = {http://dx.doi.org/10.1038/s41535-025-00814-y},
  DOI = {10.1038/s41535-025-00814-y},
  number = {1},
  journal = {npj Quantum Materials},
  publisher = {Springer Science and Business Media LLC},
  author = {León,  Andrea and Autieri,  Carmine and Brumme,  Thomas and González,  Jhon W.},
  year = {2025},
  month = sep 
}

@article{PhysRevX.15.031006,
  title = {Multipolar Anisotropy in Anomalous Hall Effect from Spin-Group Symmetry Breaking},
  author = {Liu, Zheng and Wei, Mengjie and Peng, Wenzhi and Hou, Dazhi and Gao, Yang and Niu, Qian},
  journal = {Phys. Rev. X},
  volume = {15},
  issue = {3},
  pages = {031006},
  numpages = {23},
  year = {2025},
  month = {Jul},
  publisher = {American Physical Society},
  doi = {10.1103/PhysRevX.15.031006},
  url = {https://link.aps.org/doi/10.1103/PhysRevX.15.031006}
}

@article{XIAO2025109872,
title = {TensorSymmetry: a package to get symmetry-adapted tensors disentangling spin-orbit coupling effect and establishing analytical relationship with magnetic order},
journal = {Computer Physics Communications},
pages = {109872},
year = {2025},
issn = {0010-4655},
doi = {https://doi.org/10.1016/j.cpc.2025.109872},
url = {https://www.sciencedirect.com/science/article/pii/S0010465525003741},
author = {Rui-Chun Xiao and Yuanjun Jin and Zhi-Fan Zhang and Zi-Hao Feng and Ding-Fu Shao and Mingliang Tian}
}

@article{PhysRevB.86.094413,
  title = {Noncollinear magnetism and single-ion anisotropy in multiferroic perovskites},
  author = {Weingart, Carlo and Spaldin, Nicola and Bousquet, Eric},
  journal = {Phys. Rev. B},
  volume = {86},
  issue = {9},
  pages = {094413},
  numpages = {11},
  year = {2012},
  month = {Sep},
  publisher = {American Physical Society},
  doi = {10.1103/PhysRevB.86.094413},
  url = {https://link.aps.org/doi/10.1103/PhysRevB.86.094413}
}

@article{Itoh2016,
  title = {Weyl fermions and spin dynamics of metallic ferromagnet SrRuO3},
  volume = {7},
  ISSN = {2041-1723},
  url = {http://dx.doi.org/10.1038/ncomms11788},
  DOI = {10.1038/ncomms11788},
  number = {1},
  journal = {Nature Communications},
  publisher = {Springer Science and Business Media LLC},
  author = {Itoh,  Shinichi and Endoh,  Yasuo and Yokoo,  Tetsuya and Ibuka,  Soshi and Park,  Je-Geun and Kaneko,  Yoshio and Takahashi,  Kei S. and Tokura,  Yoshinori and Nagaosa,  Naoto},
  year = {2016},
  month = jun 
}

@misc{daghofer2025altermagneticpolaronsfatealter,
      title={Altermagnetic polarons: the fate of alter magnetic band splittings at strong coupling}, 
      author={Maria Daghofer and Krzysztof Wohlfeld and Jeroen van den Brink},
      year={2025},
      eprint={2506.03261},
      archivePrefix={arXiv},
      primaryClass={cond-mat.str-el},
      url={https://arxiv.org/abs/2506.03261}, 
}

@article{https://doi.org/10.1002/aelm.202500714,
author = {Yu, Jiangtao and Li, Zezhong and Guo, Zhenzhou and Qian, Shifeng and Wang, Xiaotian and Liu, Zhuhong},
title = {Tailoring Topological States and Anomalous Transport via Magnetization Direction in MnSb},
journal = {Advanced Electronic Materials},
volume = {11},
number = {20},
pages = {e00714},
doi = {https://doi.org/10.1002/aelm.202500714},
url = {https://advanced.onlinelibrary.wiley.com/doi/abs/10.1002/aelm.202500714},
year = {2025}
}

@article{PhysRevB.84.104413,
  title = {Anomalous and planar Hall effect of orthorhombic and tetragonal SrRuO${}_{3}$ layers},
  author = {Ziese, M. and Vrejoiu, I.},
  journal = {Phys. Rev. B},
  volume = {84},
  issue = {10},
  pages = {104413},
  numpages = {8},
  year = {2011},
  month = {Sep},
  publisher = {American Physical Society},
  doi = {10.1103/PhysRevB.84.104413},
  url = {https://link.aps.org/doi/10.1103/PhysRevB.84.104413}
}

@article{Aroyo2006,
  title = {Bilbao Crystallographic Server: I. Databases and crystallographic computing programs},
  volume = {221},
  ISSN = {2194-4946},
  url = {http://dx.doi.org/10.1524/zkri.2006.221.1.15},
  DOI = {10.1524/zkri.2006.221.1.15},
  number = {1},
  journal = {Zeitschrift f\"{u}r Kristallographie - Crystalline Materials},
  publisher = {Walter de Gruyter GmbH},
  author = {Aroyo,  Mois Ilia and Perez-Mato,  Juan Manuel and Capillas,  Cesar and Kroumova,  Eli and Ivantchev,  Svetoslav and Madariaga,  Gotzon and Kirov,  Asen and Wondratschek,  Hans},
  year = {2006},
  month = jan,
  pages = {15–27}
}

@article{Nishihaya2025,
  title = {Spontaneous In‐Plane Anomalous Hall Response Observed in a Ferromagnetic Oxide},
  volume = {37},
  ISSN = {1521-4095},
  url = {http://dx.doi.org/10.1002/adma.202502624},
  DOI = {10.1002/adma.202502624},
  number = {47},
  journal = {Advanced Materials},
  publisher = {Wiley},
  author = {Nishihaya,  Shinichi and Matsuki,  Yuta and Kaminakamura,  Haruto and Sugeno,  Hiroki and Jiang,  Ming‐Chun and Murakami,  Yoshiya and Arita,  Ryotaro and Ishizuka,  Hiroaki and Uchida,  Masaki},
  year = {2025},
  month = sep 
}

@article{
doi:10.1126/sciadv.aaz8809,
author = {Libor Šmejkal  and Rafael González-Hernández  and T. Jungwirth  and J. Sinova },
title = {Crystal time-reversal symmetry breaking and spontaneous Hall effect in collinear antiferromagnets},
journal = {Science Advances},
volume = {6},
number = {23},
pages = {eaaz8809},
year = {2020},
doi = {10.1126/sciadv.aaz8809},
URL = {https://www.science.org/doi/abs/10.1126/sciadv.aaz8809},
eprint = {https://www.science.org/doi/pdf/10.1126/sciadv.aaz8809}}

@article{Smejkal22beyond,
  title = {Beyond Conventional Ferromagnetism and Antiferromagnetism: A Phase with Nonrelativistic Spin and Crystal Rotation Symmetry},
  author = {\ifmmode \check{S}\else \v{S}\fi{}mejkal, Libor and Sinova, Jairo and Jungwirth, Tomas},
  journal = {Phys. Rev. X},
  volume = {12},
  issue = {3},
  pages = {031042},
  numpages = {16},
  year = {2022},
  month = {Sep},
  publisher = {American Physical Society},
  doi = {10.1103/PhysRevX.12.031042},
  url = {https://link.aps.org/doi/10.1103/PhysRevX.12.031042}
}

@article{doi:10.1021/acsanm.9b01918,
author = {Malsch, Gerald and Ivaneyko, Dmytro and Milde, Peter and Wysocki, Lena and Yang, Lin and van Loosdrecht, Paul H. M. and Lindfors-Vrejoiu, Ionela and Eng, Lukas M.},
title = {Correlating the Nanoscale Structural, Magnetic, and Magneto-Transport Properties in SrRuO3-Based Perovskite Thin Films: Implications for Oxide Skyrmion Devices},
journal = {ACS Applied Nano Materials},
volume = {3},
number = {2},
pages = {1182-1190},
year = {2020},
doi = {10.1021/acsanm.9b01918},
URL = {https://doi.org/10.1021/acsanm.9b01918}
}

@article{Jena2021,
  title = {Interfacial Dzyaloshinskii–Moriya interaction in the epitaxial W/Co/Pt multilayers},
  volume = {13},
  ISSN = {2040-3372},
  url = {http://dx.doi.org/10.1039/D0NR08594D},
  DOI = {10.1039/d0nr08594d},
  number = {16},
  journal = {Nanoscale},
  publisher = {Royal Society of Chemistry (RSC)},
  author = {Jena,  Sukanta Kumar and Islam,  Rajibul and Milińska,  Ewelina and Jakubowski,  Marcin M. and Minikayev,  Roman and Lewińska,  Sabina and Lynnyk,  Artem and Pietruczik,  Aleksiej and Aleszkiewicz,  Paweł and Autieri,  Carmine and Wawro,  Andrzej},
  year = {2021},
  pages = {7685–7693}
}

@article{https://doi.org/10.1002/adma.202201058,
author = {Singh, Bahadur and Lin, Hsin and Bansil, Arun},
title = {Topology and Symmetry in Quantum Materials},
journal = {Advanced Materials},
volume = {35},
number = {27},
pages = {2201058},
doi = {https://doi.org/10.1002/adma.202201058},
url = {https://advanced.onlinelibrary.wiley.com/doi/abs/10.1002/adma.202201058},
year = {2023}
}

@article{6yv6-kf97,
  title = {Nature of field-induced transitions and hysteretic magnetoresistance in the noncollinear antiferromagnet ${\mathrm{EuIn}}_{2}{\mathrm{As}}_{2}$},
  author = {Singh, Karan and Skolimowski, Jan and Cuono, Giuseppe and Sattigeri, Raghottam M. and Ptok, Andrzej and Pavlosiuk, Orest and Romanova, Tetiana and Toli\ifmmode \acute{n}\else \'{n}\fi{}ski, Tomasz and Wi\ifmmode \acute{s}\else \'{s}\fi{}niewski, Piotr and Autieri, Carmine and Kaczorowski, Dariusz},
  journal = {Phys. Rev. B},
  volume = {112},
  issue = {13},
  pages = {134440},
  numpages = {11},
  year = {2025},
  month = {Oct},
  publisher = {American Physical Society},
  doi = {10.1103/6yv6-kf97},
  url = {https://link.aps.org/doi/10.1103/6yv6-kf97}
}

@article{PhysRevB.111.054442,
  title = {Staggered Dzyaloshinskii-Moriya interaction inducing weak ferromagnetism in centrosymmetric altermagnets and weak ferrimagnetism in noncentrosymmetric altermagnets},
  author = {Autieri, Carmine and Sattigeri, Raghottam M. and Cuono, Giuseppe and Fakhredine, Amar},
  journal = {Phys. Rev. B},
  volume = {111},
  issue = {5},
  pages = {054442},
  numpages = {17},
  year = {2025},
  month = {Feb},
  publisher = {American Physical Society},
  doi = {10.1103/PhysRevB.111.054442},
  url = {https://link.aps.org/doi/10.1103/PhysRevB.111.054442}
}

@article{Benny26b,
author = {Benny, M. and Gong, X.  and Fakhredine, A. and Autieri, C.},
year = {2026},
journal = {In manuscript}
}

@article{Mostofi:2008_CPC,
  title = {Wannier90: {A} Tool for Obtaining Maximally-Localised {Wannier} Functions},
  author = {Mostofi, A. A. and Yates, J. R. and Lee, Y. S. and Souza, I. and Vanderbilt, D. and Marzari, N.},
  journal = {Comput. Phys. Comm.},
  volume = {178},
  pages = {685--699},
  year = {2008},
  doi = {10.1016/j.cpc.2007.11.016},
}

@article{Smejkal22,
  title = {Emerging Research Landscape of Altermagnetism},
  author = {\ifmmode \check{S}\else \v{S}\fi{}mejkal, Libor and Sinova, Jairo and Jungwirth, Tomas},
  journal = {Phys. Rev. X},
  volume = {12},
  issue = {4},
  pages = {040501},
  numpages = {27},
  year = {2022},
  month = {Dec},
  publisher = {American Physical Society},
  doi = {10.1103/PhysRevX.12.040501},
  url = {https://link.aps.org/doi/10.1103/PhysRevX.12.040501}
}

@article{GUO2023100991,
title = {Spin-split collinear antiferromagnets: A large-scale ab-initio study},
journal = {Materials Today Physics},
volume = {32},
pages = {100991},
year = {2023},
issn = {2542-5293},
doi = {https://doi.org/10.1016/j.mtphys.2023.100991},
url = {https://www.sciencedirect.com/science/article/pii/S2542529323000275},
author = {Yaqian Guo and Hui Liu and Oleg Janson and Ion Cosma Fulga and Jeroen {van den Brink} and Jorge I. Facio}
}

@article{PhysRevB.107.155126,
  title = {Ab initio prediction of anomalous Hall effect in antiferromagnetic ${\mathrm{CaCrO}}_{3}$},
  author = {Nguyen, Thi Phuong Thao and Yamauchi, Kunihiko},
  journal = {Phys. Rev. B},
  volume = {107},
  issue = {15},
  pages = {155126},
  numpages = {9},
  year = {2023},
  month = {Apr},
  publisher = {American Physical Society},
  doi = {10.1103/PhysRevB.107.155126},
  url = {https://link.aps.org/doi/10.1103/PhysRevB.107.155126}
}

@article{Cuono23orbital,
title = {Orbital-selective altermagnetism and correlation-enhanced spin-splitting in strongly-correlated transition metal oxides},
journal = {Journal of Magnetism and Magnetic Materials},
volume = {586},
pages = {171163},
year = {2023},
issn = {0304-8853},
doi = {https://doi.org/10.1016/j.jmmm.2023.171163},
url = {https://www.sciencedirect.com/science/article/pii/S0304885323008132},
author = {Giuseppe Cuono and Raghottam M. Sattigeri and Jan Skolimowski and Carmine Autieri}
}

@article{Autieri_2016,
doi = {10.1088/0953-8984/28/42/426004},
url = {https://dx.doi.org/10.1088/0953-8984/28/42/426004},
year = {2016},
month = {sep},
publisher = {IOP Publishing},
volume = {28},
number = {42},
pages = {426004},
author = {C Autieri},
title = {Antiferromagnetic and xy ferro-orbital order in insulating SrRuO3 thin films with SrO termination},
journal = {Journal of Physics: Condensed Matter}
}

@article{Groenendijk20,
  title = {Berry phase engineering at oxide interfaces},
  author = {Groenendijk, D. J. and Autieri, C. and van Thiel, T. C. and Brzezicki, W. and Hortensius, J. R. and Afanasiev, D. and Gauquelin, N. and Barone, P. and van den Bos, K. H. W. and van Aert, S. and Verbeeck, J. and Filippetti, A. and Picozzi, S. and Cuoco, M. and Caviglia, A. D.},
  journal = {Phys. Rev. Res.},
  volume = {2},
  issue = {2},
  pages = {023404},
  numpages = {8},
  year = {2020},
  month = {Jun},
  publisher = {American Physical Society},
  doi = {10.1103/PhysRevResearch.2.023404},
  url = {https://link.aps.org/doi/10.1103/PhysRevResearch.2.023404}
}

@article{Vanthiel21,
  title = {Coupling Charge and Topological Reconstructions at Polar Oxide Interfaces},
  author = {van Thiel, T. C. and Brzezicki, W. and Autieri, C. and Hortensius, J. R. and Afanasiev, D. and Gauquelin, N. and Jannis, D. and Janssen, N. and Groenendijk, D. J. and Fatermans, J. and Van Aert, S. and Verbeeck, J. and Cuoco, M. and Caviglia, A. D.},
  journal = {Phys. Rev. Lett.},
  volume = {127},
  issue = {12},
  pages = {127202},
  numpages = {7},
  year = {2021},
  month = {Sep},
  publisher = {American Physical Society},
  doi = {10.1103/PhysRevLett.127.127202},
  url = {https://link.aps.org/doi/10.1103/PhysRevLett.127.127202}
}

@article{perdew1996generalized,
  title={Generalized gradient approximation made simple},
  author={Perdew, John P and Burke, Kieron and Ernzerhof, Matthias},
  journal={Physical review letters},
  volume={77},
  number={18},
  pages={3865},
  year={1996},
  publisher={APS}
}

@article{w90,
  title={An updated version of wannier90: A tool for obtaining maximally-localised Wannier functions},
  author={Mostofi, Arash A and Yates, Jonathan R and Pizzi, Giovanni and Lee, Young-Su and Souza, Ivo and Vanderbilt, David and Marzari, Nicola},
  journal={Computer Physics Communications},
  volume={185},
  number={8},
  pages={2309--2310},
  year={2014},
  publisher={Elsevier}
}

@article{Fang2024,
  title = {Quantum Geometry Induced Nonlinear Transport in Altermagnets},
  volume = {133},
  ISSN = {1079-7114},
  url = {http://dx.doi.org/10.1103/PhysRevLett.133.106701},
  DOI = {10.1103/physrevlett.133.106701},
  number = {10},
  journal = {Physical Review Letters},
  publisher = {American Physical Society (APS)},
  author = {Fang,  Yuan and Cano,  Jennifer and Ghorashi,  Sayed Ali Akbar},
  year = {2024},
  month = sep 
}

@article{PhysRevLett.126.127701,
  title = {Efficient Electrical Spin Splitter Based on Nonrelativistic Collinear Antiferromagnetism},
  author = {Gonz\'alez-Hern\'andez, Rafael and \ifmmode \check{S}\else \v{S}\fi{}mejkal, Libor and V\'yborn\'y, Karel and Yahagi, Yuta and Sinova, Jairo and Jungwirth, Tom\'a\ifmmode \check{s}\else \v{s}\fi{} and \ifmmode \check{Z}\else \v{Z}\fi{}elezn\'y, Jakub},
  journal = {Phys. Rev. Lett.},
  volume = {126},
  issue = {12},
  pages = {127701},
  numpages = {6},
  year = {2021},
  month = {Mar},
  publisher = {American Physical Society},
  doi = {10.1103/PhysRevLett.126.127701},
  url = {https://link.aps.org/doi/10.1103/PhysRevLett.126.127701}
}

@article{w90-ahc,
  title={Ab initio calculation of the anomalous Hall conductivity by Wannier interpolation},
  author={Wang, Xinjie and Yates, Jonathan R and Souza, Ivo and Vanderbilt, David},
  journal={Physical Review B},
  volume={74},
  number={19},
  pages={195118},
  year={2006},
  publisher={APS}
}

@article{wanniertools,
  title={WannierTools: An open-source software package for novel topological materials},
  author={Wu, QuanSheng and Zhang, ShengNan and Song, Hai-Feng and Troyer, Matthias and Soluyanov, Alexey A},
  journal={Computer Physics Communications},
  volume={224},
  pages={405--416},
  year={2018},
  publisher={Elsevier}
}

@article{PhysRevLett.130.036702,
  title = {Spontaneous Anomalous Hall Effect Arising from an Unconventional Compensated Magnetic Phase in a Semiconductor},
  author = {Gonzalez Betancourt, R. D. and Zub\'a\ifmmode \check{c}\else \v{c}\fi{}, J. and Gonzalez-Hernandez, R. and Geishendorf, K. and \ifmmode \check{S}\else \v{S}\fi{}ob\'a\ifmmode \check{n}\else \v{n}\fi{}, Z. and Springholz, G. and Olejn\'{\i}k, K. and \ifmmode \check{S}\else \v{S}\fi{}mejkal, L. and Sinova, J. and Jungwirth, T. and Goennenwein, S. T. B. and Thomas, A. and Reichlov\'a, H. and \ifmmode \check{Z}\else \v{Z}\fi{}elezn\'y, J. and Kriegner, D.},
  journal = {Phys. Rev. Lett.},
  volume = {130},
  issue = {3},
  pages = {036702},
  numpages = {7},
  year = {2023},
  month = {Jan},
  publisher = {American Physical Society},
  doi = {10.1103/PhysRevLett.130.036702},
  url = {https://link.aps.org/doi/10.1103/PhysRevLett.130.036702}
}

@article{turek2022altermagnetism,
  title={Altermagnetism and magnetic groups with pseudoscalar electron spin},
  author={Turek, Ilja},
  journal={Physical Review B},
  volume={106},
  number={9},
  pages={094432},
  year={2022},
  publisher={APS}
}

@article{shao2023neel,
  title={N{\'e}el spin currents in antiferromagnets},
  author={Shao, Ding-Fu and Jiang, Yuan-Yuan and Ding, Jun and Zhang, Shu-Hui and Wang, Zi-An and Xiao, Rui-Chun and Gurung, Gautam and Lu, WJ and Sun, YP and Tsymbal, Evgeny Y},
  journal={Physical Review Letters},
  volume={130},
  number={21},
  pages={216702},
  year={2023},
  publisher={APS}
}

@article{yuan2023degeneracy,
  title={Degeneracy Removal of Spin Bands in Collinear Antiferromagnets with Non-Interconvertible Spin-Structure Motif Pair},
  author={Yuan, Lin-Ding and Zunger, Alex},
  journal={Advanced Materials},
  pages={2211966},
  year={2023},
  publisher={Wiley Online Library}
}

@article{Brzezicki2019,
  doi = {10.1088/1361-648x/ab448d},
  url = {https://doi.org/10.1088/1361-648x/ab448d},
  year = {2019},
  month = oct,
  publisher = {{IOP} Publishing},
  volume = {32},
  number = {2},
  pages = {023001},
  author = {Wojciech Brzezicki},
  title = {Spin,  orbital and topological order in models of strongly correlated electrons},
  journal = {Journal of Physics: Condensed Matter}
}

@article{ssxp-gz9l,
  title = {Conditions for orbital-selective altermagnetism in ${\mathrm{Sr}}_{2}{\mathrm{RuO}}_{4}$: Tight-binding model, similarities with cuprates, and implications for superconductivity},
  author = {Autieri, Carmine and Cuono, Giuseppe and Chakraborty, Debmalya and Gentile, Paola and Black-Schaffer, Annica M.},
  journal = {Phys. Rev. B},
  volume = {112},
  issue = {1},
  pages = {014412},
  numpages = {20},
  year = {2025},
  month = {Jul},
  publisher = {American Physical Society},
  doi = {10.1103/ssxp-gz9l},
  url = {https://link.aps.org/doi/10.1103/ssxp-gz9l}
}

@article{10.1063/5.0158271,
    author = {Wu, A. Junxiang and Zhang, B. Zeying and Liu, C. Jian and Shao, D. Xiaohong},
    title = "{Magnetic quadratic nodal line with spin–orbital coupling in CrSb}",
    journal = {Applied Physics Letters},
    volume = {123},
    number = {5},
    pages = {052407},
    year = {2023},
    month = {08},
    issn = {0003-6951},
    doi = {10.1063/5.0158271},
    url = {https://doi.org/10.1063/5.0158271}
}

@article{Peng2025,
  title = {Unconventional scaling of the orbital Hall effect},
  volume = {24},
  ISSN = {1476-4660},
  url = {http://dx.doi.org/10.1038/s41563-025-02326-3},
  DOI = {10.1038/s41563-025-02326-3},
  number = {11},
  journal = {Nature Materials},
  publisher = {Springer Science and Business Media LLC},
  author = {Peng,  Siyang and Zheng,  Xuan and Li,  Sheng and Lao,  Bin and Han,  Yamin and Liao,  Zhaoliang and Zheng,  Hongsheng and Yang,  Yumeng and Yu,  Tianye and Liu,  Peitao and Sun,  Yan and Chen,  Xing-Qiu and Peng,  Shouzhong and Zhao,  Weisheng and Li,  Run-Wei and Wang,  Zhiming},
  year = {2025},
  month = aug,
  pages = {1749–1755}
}

@article{Brzezicki2025,
  title = {Sign Competing Sources of Berry Curvature and Anomalous Hall Conductance Humps in Topological Ferromagnets},
  volume = {11},
  ISSN = {2199-160X},
  url = {http://dx.doi.org/10.1002/aelm.202500307},
  DOI = {10.1002/aelm.202500307},
  number = {18},
  journal = {Advanced Electronic Materials},
  publisher = {Wiley},
  author = {Brzezicki,  Wojciech and Autieri,  Carmine and Cuoco,  Mario},
  year = {2025},
  month = sep 
}

@article{Kimbell2022,
  title = {Challenges in identifying chiral spin textures via the topological Hall effect},
  volume = {3},
  ISSN = {2662-4443},
  url = {http://dx.doi.org/10.1038/s43246-022-00238-2},
  DOI = {10.1038/s43246-022-00238-2},
  number = {1},
  journal = {Communications Materials},
  publisher = {Springer Science and Business Media LLC},
  author = {Kimbell,  Graham and Kim,  Changyoung and Wu,  Weida and Cuoco,  Mario and Robinson,  Jason W. A.},
  year = {2022},
  month = apr 
}

@article{Gong2026,
author = {Gong, X. and Fakhredine, A. and Autieri, C.},
year = {2026},
journal = {Submitted}
}

@article{w52v-blqm,
  title = {Anisotropic spin-polarized conductivity in collinear altermagnets},
  author = {Dou, Mingbo and Wang, Xianjie and Tao, L. L.},
  journal = {Phys. Rev. B},
  volume = {111},
  issue = {22},
  pages = {224423},
  numpages = {8},
  year = {2025},
  month = {Jun},
  publisher = {American Physical Society},
  doi = {10.1103/w52v-blqm},
  url = {https://link.aps.org/doi/10.1103/w52v-blqm}
}

@ARTICLE{Samanta2020-vj,
  title     = "Crystal Hall and crystal magneto-optical effect in thin films of
               {SrRuO3}",
  author    = "Samanta, Kartik and Le{\v z}ai{\'c}, Marjana and Merte,
               Maximilian and Freimuth, Frank and Bl{\"u}gel, Stefan and
               Mokrousov, Yuriy",
  journal   = "J. Appl. Phys.",
  publisher = "AIP Publishing",
  volume    =  127,
  number    =  21,
  pages     = "213904",
  month     =  jun,
  year      =  2020,
  language  = "en"
}

@article{PhysRevB.109.125305,
  title = {Unveiling the chirality of the quantum anomalous Hall effect},
  author = {Chen, Hongyu and Jin, Yuanjun and Yahyavi, Mohammad and Belopolski, Ilya and Shao, Sen and Hou, Tao and Wang, Naizhou and Hsu, Chia-Hsiu and Zhao, Yilin and Yang, Bo and Ma, Qiong and Yin, Jia-Xin and Xu, Su-Yang and Gao, Wei-bo and Chang, Guoqing},
  journal = {Phys. Rev. B},
  volume = {109},
  issue = {12},
  pages = {125305},
  numpages = {7},
  year = {2024},
  month = {Mar},
  publisher = {American Physical Society},
  doi = {10.1103/PhysRevB.109.125305},
  url = {https://link.aps.org/doi/10.1103/PhysRevB.109.125305}
}

@article{PhysRevLett.132.176702,
  title = {Landau Theory of Altermagnetism},
  author = {McClarty, Paul A. and Rau, Jeffrey G.},
  journal = {Phys. Rev. Lett.},
  volume = {132},
  issue = {17},
  pages = {176702},
  numpages = {8},
  year = {2024},
  month = {Apr},
  publisher = {American Physical Society},
  doi = {10.1103/PhysRevLett.132.176702},
  url = {https://link.aps.org/doi/10.1103/PhysRevLett.132.176702}
}

@article{839n-rckn,
  title = {Quasisymmetry-Constrained Spin Ferromagnetism in Altermagnets},
  author = {Roig, Merc\`e and Yu, Yue and Ekman, Rune C. and Kreisel, Andreas and Andersen, Brian M. and Agterberg, Daniel F.},
  journal = {Phys. Rev. Lett.},
  volume = {135},
  issue = {1},
  pages = {016703},
  numpages = {8},
  year = {2025},
  month = {Jul},
  publisher = {American Physical Society},
  doi = {10.1103/839n-rckn},
  url = {https://link.aps.org/doi/10.1103/839n-rckn}
}

@article{Fakhredine25b,
  title = {Interplay between Relativistic Spin-Momentum Locking and
Breaking of Inversion Symmetry},
  author = {Fakhredine, A. and Cuono, G. and Skolimowski, J. and Picozzi, S. and Autieri, C.},
  year = {2026},
  journal = {In preparation},
}

@Article{Cheong2024,
author={Cheong, Sang-Wook
and Huang, Fei-Ting},
title={Altermagnetism with non-collinear spins},
journal={npj Quantum Materials},
year={2024},
month={Jan},
day={22},
volume={9},
number={1},
pages={13},
issn={2397-4648},
doi={10.1038/s41535-024-00626-6},
url={https://doi.org/10.1038/s41535-024-00626-6}
}

@article{guo2017spin,
  title={Spin-wave canting induced by the Dzyaloshinskii-Moriya interaction in ferromagnetic nanowires},
  author={Guo, Jun and Zeng, Xiaoyan and Yan, Ming},
  journal={Physical Review B},
  volume={96},
  number={1},
  pages={014404},
  year={2017},
  publisher={APS}
}

@Article{Kipp2021,
author={Kipp, Jonathan
and Samanta, Kartik
and Lux, Fabian R.
and Merte, Maximilian
and Go, Dongwook
and Hanke, Jan-Philipp
and Redies, Matthias
and Freimuth, Frank
and Bl{\"u}gel, Stefan
and Le{\v{z}}ai{\'{c}}, Marjana
and Mokrousov, Yuriy},
title={The chiral Hall effect in canted ferromagnets and antiferromagnets},
journal={Communications Physics},
year={2021},
month={May},
day={14},
volume={4},
number={1},
pages={99},
issn={2399-3650},
doi={10.1038/s42005-021-00587-3},
url={https://doi.org/10.1038/s42005-021-00587-3}
}

@article{PhysRev.120.91,
  title = {Anisotropic Superexchange Interaction and Weak Ferromagnetism},
  author = {Moriya, T\^oru},
  journal = {Phys. Rev.},
  volume = {120},
  issue = {1},
  pages = {91--98},
  numpages = {0},
  year = {1960},
  month = {Oct},
  publisher = {American Physical Society},
  doi = {10.1103/PhysRev.120.91},
  url = {https://link.aps.org/doi/10.1103/PhysRev.120.91}
}

@article{PhysRevB.105.245107,
  title = {Role of spin-orbit coupling in canted ferromagnetism and spin-wave dynamics of ${\mathrm{SrRuO}}_{3}$},
  author = {Ahn, Kyo-Hoon and Marmodoro, Alberto and Hejtmánek, Jiří and Jirák, Zdeněk and Knížek, Karel},
  journal = {Phys. Rev. B},
  volume = {105},
  issue = {24},
  pages = {245107},
  numpages = {8},
  year = {2022},
  month = {Jun},
  publisher = {American Physical Society},
  doi = {10.1103/PhysRevB.105.245107},
  url = {https://link.aps.org/doi/10.1103/PhysRevB.105.245107}
}

@article{PhysRevLett.124.067203,
  title = {Nonlinear Anomalous Hall Effect for N\'eel Vector Detection},
  author = {Shao, Ding-Fu and Zhang, Shu-Hui and Gurung, Gautam and Yang, Wen and Tsymbal, Evgeny Y.},
  journal = {Phys. Rev. Lett.},
  volume = {124},
  issue = {6},
  pages = {067203},
  numpages = {7},
  year = {2020},
  month = {Feb},
  publisher = {American Physical Society},
  doi = {10.1103/PhysRevLett.124.067203},
  url = {https://link.aps.org/doi/10.1103/PhysRevLett.124.067203}
}

@article{10.1063/5.0100912,
    author = {Cuoco, M. and Di Bernardo, A.},
    title = {Materials challenges for SrRuO3: From conventional to quantum electronics},
    journal = {APL Materials},
    volume = {10},
    number = {9},
    pages = {090902},
    year = {2022},
    month = {09},
    issn = {2166-532X},
    doi = {10.1063/5.0100912},
    url = {https://doi.org/10.1063/5.0100912}
}

@article{10.1063/5.0252836,
    author = {Bernardini, Fabio and Fiebig, Manfred and Cano, Andrés},
    title = {Ruddlesden–Popper and perovskite phases as a material platform for altermagnetism},
    journal = {Journal of Applied Physics},
    volume = {137},
    number = {10},
    pages = {103903},
    year = {2025},
    month = {03},
    issn = {0021-8979},
    doi = {10.1063/5.0252836},
    url = {https://doi.org/10.1063/5.0252836}
}

@Article{D1MH01385H,
author ="Li, Zhe and Chen, Xiaobing and Chen, Yuansha and Zhang, Qinghua and Zhang, Hui and Zhang, Jine and Shi, Wenxiao and He, Bin and Zhang, Jinxing and Song, Jinghua and Han, Furong and Liu, Banggui and Gu, Lin and Hu, Fengxia and Chen, Yunzhong and Shen, Baogen and Sun, Jirong",
title  ="Infinite-layer/perovskite oxide heterostructure-induced high-spin states in SrCuO2/SrRuO3 bilayer films",
journal  ="Mater. Horiz.",
year  ="2021",
volume  ="8",
issue  ="12",
pages  ="3468-3476",
publisher  ="The Royal Society of Chemistry",
doi  ="10.1039/D1MH01385H",
url  ="http://dx.doi.org/10.1039/D1MH01385H"}

@misc{lu2025insulatortometaltransitionmagneticreconstruction,
      title={Insulator-to-Metal Transition via Magnetic Reconstruction at Oxide Interfaces}, 
      author={Zengxing Lu and Jiatai Feng and Xuan Zheng and You-guo Shi and Run-Wei Li and Carmine Autieri and Mario Cuoco and Milan Radovic and Zhiming Wang},
      year={2025},
      eprint={2503.21093},
      archivePrefix={arXiv},
      primaryClass={cond-mat.mes-hall},
      url={https://arxiv.org/abs/2503.21093}, 
}

@misc{yoo2025micromagneticformalismmagneticmultipoles,
      title={Micromagnetic formalism for magnetic multipoles}, 
      author={Myoung-Woo Yoo and Axel Hoffmann},
      year={2025},
      eprint={2501.07513},
      archivePrefix={arXiv},
      primaryClass={cond-mat.mes-hall},
      url={https://arxiv.org/abs/2501.07513}, 
}

@misc{jskol_SOC_Code_V1_2025,
  author       = {Skolimowski, Jan and Autieri, Carmine and Jamroszczyk, Kamil and Benny, Mathews},
  title        = {SOC\_Code\_V1},
  year         = {2025},
  howpublished = {\url{https://github.com/jskol/SOC_Code_V1}},
  note         = {Accessed: 2025-11-11}
}

@article{Derunova2025,
  title = {A Fermi surface descriptor quantifying the correlations between anomalous Hall effect and Fermi surface geometry},
  volume = {8},
  ISSN = {2666-9366},
  url = {http://dx.doi.org/10.21468/SciPostPhysCore.8.4.085},
  DOI = {10.21468/scipostphyscore.8.4.085},
  number = {4},
  journal = {SciPost Physics Core},
  publisher = {Stichting SciPost},
  author = {Derunova,  Elena and Gayles,  Jacob and Sun,  Yan and Gaultois,  Michael W. and Ali,  Mazhar N.},
  year = {2025},
  month = nov 
}

@article{PhysRevB.91.205116,
  title = {Nature of itinerant ferromagnetism of ${\mathrm{SrRuO}}_{3}$: A DFT+DMFT study},
  author = {Kim, Minjae and Min, B. I.},
  journal = {Phys. Rev. B},
  volume = {91},
  issue = {20},
  pages = {205116},
  numpages = {6},
  year = {2015},
  month = {May},
  publisher = {American Physical Society},
  doi = {10.1103/PhysRevB.91.205116},
  url = {https://link.aps.org/doi/10.1103/PhysRevB.91.205116}
}

@article{10.1063/5.0043057,
    author = {Piamonteze, Cinthia and Bern, Francis and Avula, Sridhar Reddy Venkata and Studniarek, Michał and Autieri, Carmine and Ziese, Michael and Lindfors-Vrejoiu, Ionela},
    title = {Ferromagnetic order of ultra-thin La0.7Ba0.3MnO3 sandwiched between SrRuO3 layers},
    journal = {Applied Physics Letters},
    volume = {118},
    number = {15},
    pages = {152408},
    year = {2021},
    month = {04},
    issn = {0003-6951},
    doi = {10.1063/5.0043057},
    url = {https://doi.org/10.1063/5.0043057}
}

@article{Tian2021,
  title = {Manipulating Berry curvature of SrRuO3thin films via epitaxial strain},
  volume = {118},
  ISSN = {1091-6490},
  url = {http://dx.doi.org/10.1073/pnas.2101946118},
  DOI = {10.1073/pnas.2101946118},
  number = {18},
  journal = {Proceedings of the National Academy of Sciences},
  publisher = {Proceedings of the National Academy of Sciences},
  author = {Tian,  Di and Liu,  Zhiwei and Shen,  Shengchun and Li,  Zhuolu and Zhou,  Yu and Liu,  Hongquan and Chen,  Hanghui and Yu,  Pu},
  year = {2021},
  month = apr 
}

@article{Lu2022,
  title = {Cooperative control of perpendicular magnetic anisotropy via crystal structure and orientation in freestanding SrRuO3 membranes},
  volume = {6},
  ISSN = {2397-4621},
  url = {http://dx.doi.org/10.1038/s41528-022-00141-3},
  DOI = {10.1038/s41528-022-00141-3},
  number = {1},
  journal = {npj Flexible Electronics},
  publisher = {Springer Science and Business Media LLC},
  author = {Lu,  Zengxing and Yang,  Yongjie and Wen,  Lijie and Feng,  Jiatai and Lao,  Bin and Zheng,  Xuan and Li,  Sheng and Zhao,  Kenan and Cao,  Bingshan and Ren,  Zeliang and Song,  Dongsheng and Du,  Haifeng and Guo,  Yuanyuan and Zhong,  Zhicheng and Hao,  Xianfeng and Wang,  Zhiming and Li,  Run-Wei},
  year = {2022},
  month = feb 
}

@article{Huang2021,
  title = {First-Principles Calculations Predict Tunable Large Magnetic Anisotropy Due to Spin-Polarized Quantum-Well Resonances in Nanometer-Thick SrRuO3 Films: Implications for Spintronic Devices},
  volume = {4},
  ISSN = {2574-0970},
  url = {http://dx.doi.org/10.1021/acsanm.1c00775},
  DOI = {10.1021/acsanm.1c00775},
  number = {6},
  journal = {ACS Applied Nano Materials},
  publisher = {American Chemical Society (ACS)},
  author = {Huang,  Angus and Jeng,  Horng-Tay and Chang,  Ching-Hao},
  year = {2021},
  month = may,
  pages = {5932–5939}
}

@article{Filipoiu2024,
  title = {First principles electron transport in magnetoelectric SrRuO3/BaTiO3/SrTiO3/SrRuO3 interfaces},
  volume = {36},
  ISSN = {1361-6528},
  url = {http://dx.doi.org/10.1088/1361-6528/ad960f},
  DOI = {10.1088/1361-6528/ad960f},
  number = {7},
  journal = {Nanotechnology},
  publisher = {IOP Publishing},
  author = {Filipoiu,  Nicolae and Plugaru,  Neculai and Sandu,  Titus and Plugaru,  Rodica and Alexandru Nemnes,  George},
  year = {2024},
  month = dec,
  pages = {075702}
}

@article{PhysRevB.90.125109,
  title = {Strain-induced metal-insulator transition in ultrathin films of ${\text{SrRuO}}_{3}$},
  author = {Gupta, Kapil and Mandal, Basudeb and Mahadevan, Priya},
  journal = {Phys. Rev. B},
  volume = {90},
  issue = {12},
  pages = {125109},
  numpages = {7},
  year = {2014},
  month = {Sep},
  publisher = {American Physical Society},
  doi = {10.1103/PhysRevB.90.125109},
  url = {https://link.aps.org/doi/10.1103/PhysRevB.90.125109}
}

@article{PhysRevB.90.165130,
  title = {Electronic structure, cohesive properties, and magnetism of ${\mathrm{SrRuO}}_{3}$},
  author = {Gr\aa{}n\"as, Oscar and Di Marco, Igor and Eriksson, Olle and Nordstr\"om, Lars and Etz, Corina},
  journal = {Phys. Rev. B},
  volume = {90},
  issue = {16},
  pages = {165130},
  numpages = {11},
  year = {2014},
  month = {Oct},
  publisher = {American Physical Society},
  doi = {10.1103/PhysRevB.90.165130},
  url = {https://link.aps.org/doi/10.1103/PhysRevB.90.165130}
}

@article{Fakhredine26,
author = {Autieri, Carmine and Fakhredine, Amar},
title = {Relativistic Spin-Momentum Locking in Altermagnets},
journal = {The Journal of Physical Chemistry Letters},
volume = {17},
number = {2},
pages = {449-455},
year = {2026},
doi = {10.1021/acs.jpclett.5c03677},
    note ={PMID: 41469325},
URL = { https://doi.org/10.1021/acs.jpclett.5c03677}
}

@article{PhysRevB.77.214410,
  title = {Manipulating magnetic properties of ${\text{SrRuO}}_{3}$ and ${\text{CaRuO}}_{3}$ with epitaxial and uniaxial strains},
  author = {Zayak, A. T. and Huang, X. and Neaton, J. B. and Rabe, Karin M.},
  journal = {Phys. Rev. B},
  volume = {77},
  issue = {21},
  pages = {214410},
  numpages = {6},
  year = {2008},
  month = {Jun},
  publisher = {American Physical Society},
  doi = {10.1103/PhysRevB.77.214410},
  url = {https://link.aps.org/doi/10.1103/PhysRevB.77.214410}
}

@article{RevModPhys.82.1539,
  title = {Anomalous Hall effect},
  author = {Nagaosa, Naoto and Sinova, Jairo and Onoda, Shigeki and MacDonald, A. H. and Ong, N. P.},
  journal = {Rev. Mod. Phys.},
  volume = {82},
  issue = {2},
  pages = {1539--1592},
  numpages = {0},
  year = {2010},
  month = {May},
  publisher = {American Physical Society},
  doi = {10.1103/RevModPhys.82.1539},
  url = {https://link.aps.org/doi/10.1103/RevModPhys.82.1539}
}

@article{dosSantosDias2023,
  title = {Topological magnons driven by the Dzyaloshinskii-Moriya interaction in the centrosymmetric ferromagnet Mn5Ge3},
  volume = {14},
  ISSN = {2041-1723},
  url = {http://dx.doi.org/10.1038/s41467-023-43042-3},
  DOI = {10.1038/s41467-023-43042-3},
  number = {1},
  journal = {Nature Communications},
  publisher = {Springer Science and Business Media LLC},
  author = {dos Santos Dias,  M. and Biniskos,  N. and dos Santos,  F. J. and Schmalzl,  K. and Persson,  J. and Bourdarot,  F. and Marzari,  N. and Bl\"{u}gel,  S. and Br\"{u}ckel,  T. and Lounis,  S.},
  year = {2023},
  month = nov 
}
\end{document}